# Ag(II) as Spin Super-Polarizer in Molecular Spin Clusters

*Mateusz Domański[a], Jan van Leusen[b], Marvin Metzelaars,[b] Paul Kögerler*[b], and Wojciech Grochala*[a]*

[a] Mateusz Domański, Wojciech Grochala,* Center of New Technologies, University of Warsaw, Zwirki i Wigury 93, 02089 Warsaw Poland, e-mail: w.grochala@cent.uw.edu.pl; [b] Jan van Leusen, Marvin Metzelaars, Paul Kögerler,* Institut für Anorganische Chemie, RWTH Aachen University, 52056 Aachen, Germany, e-mail: paul.koegerler@ac.rwth-aachen.de

ABSTRACT Using quantum mechanical calculations, we examine magnetic (super)exchange interactions in hypothetical, chemically reasonable molecular coordination clusters containing fluoride-bridged late transition metals or selected lanthanides, as well as Ag(II). By referencing to analogous species comprising closed-shell Cd(II) we provide theoretical evidence that the presence of Ag(II) may modify the magnetic properties of such systems (including metal-metal superexchange) to a surprising degree, specifically both coupling sign and strength may markedly change. Remarkably, this happens in spite of the fact that the fluoride ligand is the least susceptible to spin polarization among all monoatomic ligands known in chemistry. In an extreme case of a oxo-bridged $Ni(II)_2$ complex, the presence of Ag(II) leads to a nearly 17-fold increase of magnetic superexchange and switching from antiferro- to ferromagnetic coupling. Ag(II) – with one hole in its d shell that may be shared with or transferred to ligands – effectively acts as spin super-polarizer, and this feature could be exploited in spintronics and diverse molecular devices.

## INTRODUCTION

Spintronics is one of the vividly developing fields of modern science and technology.[1–7] This vast subdiscipline of nanotechnology, drawing from physics, chemistry, and materials science, relies on the manipulation of individual spins either in the solid state or in small molecular assemblies using light, electric and/or magnetic fields. In this context, controllable and precisely engineered molecular devices are expected to revolutionize information processing in terms of density, speed and energy consumption. Realizing key spintronics functionalities on a molecular scale commonly mandates high magnetic anisotropy and robust coupling, mirroring the merits that have been strived for over the past two decades in the development of single-molecule magnets (SMMs) that feature either transition metals (TM, d-block elements), lanthanides (Ln, f-block elements) or both.[8–13] While Ln-based polynuclear clusters may exhibit many high multiplet states and pronounced anisotropy, the inherently weak[14] Ln–Ln magnetic interactions limit the prospects for exchange-coupled high-spin systems. On the other hand, clusters containing TMs can assume fewer quantum states but the characteristics of their magnetically relevant orbitals lead to significantly higher coupling energies. One compromise at hand is to construct polynuclear clusters containing both TM and Ln centers. Still, there is a pressing need to manipulate and greatly increase the magnetic coupling interactions present in such systems.[15–22]

Two families of compounds stand out among the materials exhibiting the strongest magnetic interactions known to date: oxocuprates(II) and fluoroargentates(II) (see **Table 1**). Magnetic superexchange interactions via a single O (or, respectively, F) bridge range up to –260 meV (–2097 $cm^{-1}$) in the former and up to –331 meV (–2670 $cm^{-1}$) for the latter, but they are consistently strong in quasi-molecular (0D), one-dimensional and two-dimensional systems. Substantial mixing of the metal d with the nonmetal p valence states ("covalence") is one reason for these materials' uniqueness.[23–27]

It might be anticipated that these two late d-shell metal cations may induce a strong electron correlation in systems comprising TM or Ln cations. Indeed, copper clusters were already extensively studied in the context of mediating or inducing magnetic exchange with lanthanide cations, focusing on Cu-Ln exchange[28–30]. Such Cu-Ln interaction usually is much stronger than pure Ln-Ln one, which even for a short distance and Ln-($\mu_2$-O)$_2$-Ln bridge is well below 1 $cm^{-1}$, e.g. –0.178 $cm^{-1}$ for a typical Dy(III) complex[31]. Usually, the coupling energy $J$ associated with a Ln-($\mu_2$-O)$_2$-Cu bridge is ferromagnetic, ranging from ~0 up to +13 $cm^{-1}$.[28–30] The corresponding Ag(II) systems have not yet been studied in experiment or theoretically. Nonetheless, Ag(II) is exceptional among the group 11 element cations, as here spin polarization may strongly affect even the attached fluoride anion. Although this ligand is not capable of transmitting strong magnetic superexchange[32–34] (as fluorine is the least spin-polarizable monoatomic ligand species known in chemistry), the presence of Ag(II) effectively enforces such an interaction[35] even in fluorides. Thus, the possibility of introducing Ag(II) to lanthanide or TM-based fluoride core structures as single-molecular magnet moieties is enticing.

**Table 1.** Selection of the compounds featuring giant absolute values of the magnetic superexchange constant between two *identical* metal centers, quoting the largest values given in the literature for each compound. Magnetic dimensionality indicates in how many directions the said interaction is operational.

| Formula | Superexchange pathway | $J$ / meV [a] | Magnetic dimensionality | Reference |
|---|---|---|---|---|
| $La_2CuO_4$ | Cu–O–Cu | –153 | 2D | 36 |
| $Sr_2CuO_2Cl_2$ | Cu–O–Cu | –130 | 2D | 37 |
| $Sr_2CuO_3$ | Cu–O–Cu | –260 | 1D | 38 |
| $AgF_2$ | Ag–F–Ag | –70 | 2D | 39 |
| [$AgF_2$][b] | Ag–F–Ag | –265[d] | 2D | 40 |
| HT-$KAgF_3$ | Ag–F–Ag | –100 | 1D | 41 |
| LT-$KAgF_3$ | Ag–F–Ag | –113[d] | 1D | 42,43 |
| $RbAgF_3$ | Ag–F–Ag | –144[d] | 1D | 42,43 |
| $CsAgF_3$ | Ag–F–Ag | –161[d] | 1D | 42,43 |
| $AgFBF_4$ | Ag–F–Ag | –331[d] | 1D | 42,43 |
| $Ag_2ZnZr_2F_{14}$ | Ag–F–Ag | –313[d] | 0D | 44 |
| $AgF_2$-HP2[c] | Ag–F–Ag | –250[d] | 0D | 44 |
| $Rb_3Ag_2F_7$ | Ag–F–Ag | –240[d] | 0D | 45 |

[a] Using a Heisenberg-Dirac-van Vleck Hamiltonian for two interacting spin centers of the type $H = -\sum_{ij} J_{ij} s_i s_j$ with ($J_{ij} = J_{ji}$); consequently, a negative $J$ value indicates antiferromagnetic interaction. [b] Single layer deposited on $RbMgF_3$ substrate. [c] High pressure structure type 2. [d] Theoretical value.

For instance, in ferromagnetic systems containing $AgF_4^{2-}$ anions, as much as half of the whole spin density may be located on four fluoride ligands, while the remainder resides on the Ag(II) center.[46] Such unusual properties allow us to treat both Cu(II) oxides and Ag(II) fluorides as redox non-innocent ligand systems. This motivates using Ag(II) as a spin polarizer toward a nonmetal ligand, which is bridging two other magnetic metals (TM or Ln) and alters their magnetic superexchange in this way. This should be a particularly promising approach if the said bridging ligand could carry a net spin and as such would constitute an equivalent of an organic free radical. We note that superexchange is well known to become very strong if the bridging ligand has free radical character.[47–49] Apart from fluoride systems, which benefit from thermodynamic stability in the case of Ag(II),[25] we can also envisage the presence of Ag(II) in an oxide system. Ag(II) compounds bearing monoatomic oxide anions are not known, but those with more complex oxoanions (sulfates, fluorosulfates, triflates, difluorophosphonates etc.) have been prepared. One argument for looking at oxide systems is that the spin-spin interactions might be transferred even easier than for fluoride, due to more substantial spin polarization of the ligand. A similar concept could be extended to other anions more prone to oxidation than fluoride, e.g. chloride, thus effectively increasing the ligand's spin and strengthening the superexchange coupling.

Herein, we theoretically study the influence of the strong spin-polarizer, Ag(II), on magnetic interactions of diverse open-shell cations' including late TMs and Lns. Specifically, we examine magnetic (super)exchange interactions in hypothetical, yet chemically reasonable molecular coordination clusters containing fluoride-bridged bimetallic systems and Ag(II). By referencing analogous species comprising closed-shell Cd(II) we provide theoretical evidence that the presence of Ag(II) may modify the magnetic properties of such systems (including metal-metal superexchange) to a marked degree, where both coupling sign and strength may change.

## METHODS

Recently numerous computational studies have centered on exchange coupling in molecular systems containing one or more lanthanide (Ln) cations, e.g. cerium,[50] praseodymium,[51] europium,[31] gadolinium,[31] terbium,[31,52] dysprosium,[31,53] or holmium[31,53]. The pronounced effort in theoretical research on magnetic clusters containing lanthanides is connected to a correct description of 4f semi-core electrons. These levels are subject to a splitting originating from ligand (crystal) field and spin-orbit interactions.[54] In many of these studies[31,50–53] the spin-orbit coupling effects have been included. However, for most gadolinium studies[29,55–61] where the spin-orbit coupling effect is predicted to be weak, the scalar-relativistic methods and the Ising-type Hamiltonian were used. One important issue with spin-orbit coupling is that it can be relatively easy employed for single-center systems, and usually a CAS approach is used for this. However, for larger systems, this approach becomes often too expensive, though the necessity persists to correctly account for relativistic effects.[29] Some authors have suggested that in Ln systems, much more important than including spin-orbit coupling is assuring the correct ground state electronic configuration (which otherwise could induce differential correlation effects and would not systematically improve the calculated results).[62]

Due to the extensive range of hypothetical molecules considered in this study, we chose an approach of hybrid DFT approach using the B3LYP functional[63,64] that is frequently used in similar studies, together with a scalar relativistic Hamiltonian ZORA and all-electron basis set SARC-ZORA-TZVP for Ag and 4f metals, and ZORA-def2-TZVP basis for the lighter elements.[62,65] The exchange spin coupling parameters $J$ were obtained using the "broken symmetry" (BS) formalism[66] by calculating all possible spin states in each system and solving the set of linear equations. For this type of molecules, it was shown that hybrid functionals may overestimate the experimentally obtained coupling constants, however to a small degree.[29]

To validate this methodology, we have cross-checked our calculations with several examples from the literature where either experimental or high-level theoretical parameters are known. We have obtained satisfying results for both Cu-Ln and Ln-Ln superexchange interactions (see **SI** and **Table S1**). As we experienced, differences in electron orbital configurations of cations in excited states would induce enormous errors for magnetic superexchange interactions, thus a great effort has been put into assuring that this is not the case here (see **Tables S7-S12**). More details of the computational methods are presented in the **SI**. Importantly, since our study focuses on a comparison between Ag(II) and related Cd(II) systems, a large share of methodological errors cancels out, thus showcasing the important trends more clearly. The use of Cd(II) systems as reference is dictated by similarity of ionic radii between Ag(II) and Cd(II) as well as the fact that Cd(II) is a closed-shell cation which does not permit facile transmission of the magnetic superexchange either.

## RESULTS

We have tested the above-mentioned overarching idea on triangular model clusters comprising a single Ag(II) and two other metal cations (**Figure 1**). This particular geometry type was chosen as it facilitates the influence of Ag(II) directly on the Ln-X-Ln bridge, which is the purpose of this work. Such bridges are common in the solid state, where a ligand bridging two larger cations is also coordinated to a smaller one, as present in e.g. the crystalline lattices of $La_2CuO_4$ or the K-Ag(II)-F series. Concurrently, the cluster geometry derives partly from the postulated $Ln_2F_6$ molecules[67–69] and bridges present in some larger molecular magnets with Ln-($\mu_2$-X)$_2$-Ln bridges.[28–31] Here,[32–34] the $AgF^+$ group is added, in a perpendicular orientation, to the center of the Ln-X-Ln bridge with other fluorides mainly stabilizing the central part. The chosen molecular clusters proved their stability in numerous geometry optimization runs, and as such serve as testbed for the spin-spin interactions.[35]

We note that the employed clusters including quasi-triangular trimetallic core may be subject to geometrical spin-frustration. In particular, the antiferromagnetic interactions between two metallic centers and Ag(II) may compete with the antiferromagnetic coupling between two metal cations. Nevertheless, as will be shown below, most often the Ag-M interactions predominate and they determine the spin ordering of the ground state.

Following our recent study we have chosen the three late 3d TMs Cu(II), high-spin (HS) Ni(II) and HS Co(II), since they offer a broad variety of spin moments from spin-½ to spin-$\frac{3}{2}$, and of metal-ligand bond covalence. Simultaneously, we have avoided using early TMs, which would be immediately oxidized by the Ag(II) cation.[25] The bridging ligands, X, comprise a fluoride $F^-$ (a ligand of choice in electron-deficient Ag(II) systems) and its isoelectronic siblings, $O^{2-}$ and $Cl^-$. For each M, the cluster geometry was optimized in its electronic ground state configuration. To understand the impact of Ag(II) on the magnetic properties of the cluster, and in particular, on the M–M magnetic superexchange, we have referenced the results to those obtained for isostructural derivatives containing Cd instead of Ag. Cd(II) is a closed-shell diamagnetic cation incapable of transmitting magnetic superexchange to any significant degree and, therefore, the magnetic properties of Cd-bearing species are dominated by the contributions of the M–M superexchange mediated by the X bridge. Since the ionic radius of Cd(II) (Shannon-Prewitt provide the value of 1.09 Å for octahedral coordination) is very close to that of Ag(II) (1.08 Å for octahedral coordination), and in order to extract the impact of Ag(II) on electronic and magnetic properties free from strain effects, we kept the molecular geometry of each Cd-substituted species identical to that of its Ag(II) analogue.

In a first step, we explored the three afore-mentioned 3d TM cations and the entire series of partially filled f-subshell lanthanide cations in the given cluster geometry with the three central ligands, X: fluoride, and chloride or oxide as alternatives. We notice that, in some cases, a reorganization of the cluster geometry occurred, in particular with 3d TM cations, while some lanthanide clusters did not yield stable arrangement. In other cases, a spontaneous redox reaction took place. For 3d TM cation clusters, the electronic and ionic stability was obtained for systems with fluoride bridging ligands for all cations, with chlorides for Co and Ni, and with oxide only for Ni cations. In the case of lanthanide cations clusters, we obtained electronic and ionic stability with the fluoride bridging ligand for $f^3$ – $f^5$ cations, with chlorides for $f^2$ – $f^{10}$ cations, and with oxide for $f^5$ – $f^{13}$ cations. We here discuss all cases where the geometry scheme shown in **Figure 1** represents a stable local minimum, only in two cases being associated with an excited spin state. The obtained changes of metal-metal superexchange interaction due to Ag(II) influence are presented in **Figure 2**, and numerical data are gathered in **Table 2** (for further details see **Figures S1**, **S2** and **Table S2** in the **SI**). The resulting spin populations of the key atoms in the studied clusters are listed in **Table S4**.

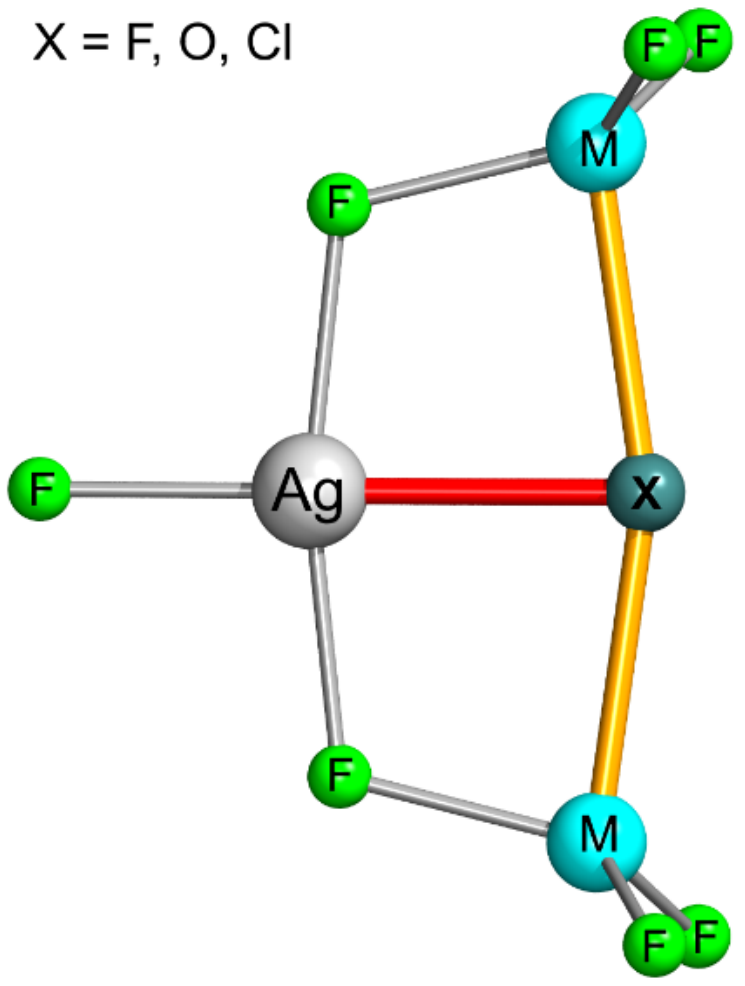

**Figure 1.** Schematic drawing of all optimized geometries of small metal-silver coordination clusters studied in this work. M–(μ-X)–M bonds are highlighted in orange, Ag–X bond in red. The reference systems feature $d^{10}$ $Cd^{2+}$ instead of $d^9$ $Ag^{2+}$. The stoichiometries are anionic $[M_2AgF_8]^{2-}$ (M

= Co, Ni, Cu), $[M_2AgF_7Cl]^{2-}$ (M = Co, Ni), $[M_2AgF_7O]^{3-}$ (M = Co, Ni, Cu), $[M_2AgF_7O]^{-}$ (M = Sm, Eu, Gd, Tb, Dy, Ho, Er, Tm, Yb), neutral $M_2AgF_8$ (M = Nd, Pm, Sm) and $M_2AgF_7Cl$ (M = Pr, Nd, Pm, Sm, Eu, Gd, Tb, Dy, Ho). M is divalent for Co, Ni and Co, and trivalent for lanthanides.

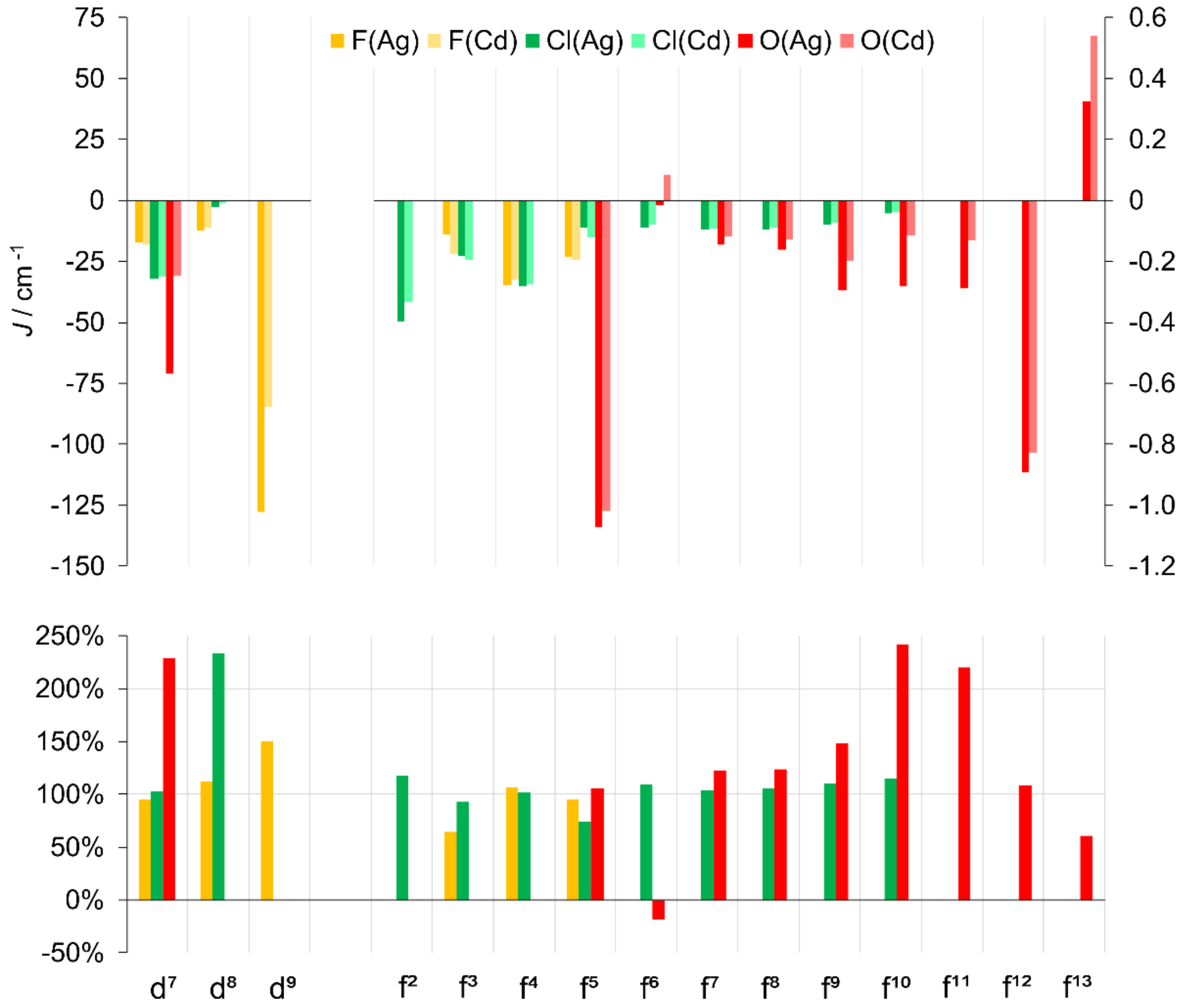

**Figure 2.** Impact of Ag(II) on metal-metal superexchange constant, $J_{MM}$, as measured by (top) the values of $J_{MM}$, and (bottom) the ratio (in %) of $J_{MM}$ in the Ag(II)- vs. Cd(II)-containing species, free from magnetostrictive effects. Here, +100% corresponds to no impact, whereas negative values imply a change in the interaction from AFM to FM or *vice versa*. *J* scales for TMs and Lns are separate. The results for the $Ni_2$ cluster with central oxide ligands (cf. **Table 2**) are off the scale and have not been shown here.

**Table 2.** Computed data for each considered system: valence electron configuration (v.e.), ground state configuration (GS), values of silver-metal and metal-metal superexchange constants before and after Ag(II) → Cd(II) substitution ($J_{AgM}$, $J_{MM}$(Ag) and $J_{MM}$(Cd), respectively) and brief analysis of these values. Each system has one GS out of three possible spin-states: high-spin "HS", broken-symmetry with S(Ag) of opposite polarity "BS(Ag)", or broken-symmetry with S(M) of opposite polarity "BS(M)". The ratio is defined as Ratio = $J_{MM}$(Ag)/$J_{MM}$(Cd), while the difference as Δ = $J_{MM}$(Ag)–$J_{MM}$(Cd). Clusters in principle may lack symmetry elements thus $J_{AgM}$ may be an average

over two interactions (Ag–M[1] and Ag–M[2]), *cf.* **SI** for details. Asterisks indicate that an electronic configuration, in which cluster geometry is stable, becomes metastable with respect to the other electronic configuration when relaxed (see more in **SI**); however, such cases are rare.

| Bridging Ligand | Metal | v.e. | GS | $J_{AgM}$ / $cm^{-1}$ | $J_{MM}$(Ag) / $cm^{-1}$ | $J_{MM}$(Cd) / $cm^{-1}$ | Ratio | Δ / $cm^{-1}$ |
|---|---|---|---|---|---|---|---|---|
| $F^-$ | Co | $d^7$ | BS(M) | 41.0 | –17.3 | –18.2 | 95% | 0.9 |
| | Ni | $d^8$ | BS(Ag) | –44.4 | –12.5 | –11.1 | 112% | –1.4 |
| | Cu | $d^9$ | BS(Ag) | –285.5 | –127.6 | –84.7 | 151% | –42.9 |
| | Nd | $f^3$ | BS(M) | –9.49 | –0.11 | –0.17 | 64% | 0.063 |
| | Pm | $f^4$ | BS(Ag) | –1.78 | –0.28 | –0.26 | 106% | –0.016 |
| | Sm | $f^5$ | HS | 2.69 | –0.18 | –0.19 | 95% | 0.009 |
| $O^{2-}$ | Co | $d^7$ | HS* | 326.5 | –71.1 | –31.1 | 229% | –40.0 |
| | Ni | $d^8$ | HS | 409.1 | 1670.6 | –100.5 | –1662% | 1771.1 |
| | Cu | $d^9$ | BS(M)* | –255.2 | –608.7 | –381.0 | 160% | –227.7 |
| | Sm | $f^5$ | BS(Ag) | –29.94 | –1.073 | –1.018 | 105% | –0.055 |
| | Eu | $f^6$ | BS(Ag) | –22.60 | –0.016 | 0.084 | –19% | –0.100 |
| | Gd | $f^7$ | BS(Ag) | –12.54 | –0.146 | –0.119 | 123% | –0.027 |
| | Tb | $f^8$ | BS(M) | –4.30 | –0.160 | –0.130 | 123% | –0.030 |
| | Dy | $f^9$ | BS(M) | –1.47 | –0.293 | –0.197 | 148% | –0.096 |
| | Ho | $f^{10}$ | HS | 2.85 | –0.281 | –0.116 | 242% | –0.164 |
| | Er | $f^{11}$ | BS(Ag) | –8.99 | –0.289 | –0.131 | 220% | –0.157 |
| | Tm | $f^{12}$ | HS | 21.71 | –0.893 | –0.826 | 108% | –0.067 |
| | Yb | $f^{13}$ | HS | 19.61 | 0.326 | 0.540 | 60% | –0.214 |
| $Cl^-$ | Co | $d^7$ | BS(M) | 7.7 | –32.3 | –31.5 | 103% | –0.8 |
| | Ni | $d^8$ | BS(Ag) | –83.1 | –2.7 | –1.1 | 234% | –1.5 |
| | Pr | $f^2$ | BS(Ag) | –31.44 | –0.395 | –0.334 | 118% | –0.060 |
| | Nd | $f^3$ | BS(M) | –4.48 | –0.183 | –0.196 | 93% | 0.013 |

| | | | | | | | |
|---|---|---|---|---|---|---|---|
| Pm | $f^4$ | BS(Ag) | –1.95 | –0.279 | –0.275 | 101% | –0.004 |
| Sm | $f^5$ | BS(M) | 2.11 | –0.090 | –0.122 | 74% | 0.031 |
| Eu | $f^6$ | HS | 2.91 | –0.087 | –0.080 | 109% | –0.007 |
| Gd | $f^7$ | HS | 2.11 | –0.096 | –0.092 | 104% | –0.003 |
| Tb | $f^8$ | HS | 4.93 | –0.095 | –0.090 | 105% | –0.005 |
| Dy | $f^9$ | HS | 6.88 | –0.078 | –0.071 | 111% | –0.007 |
| Ho | $f^{10}$ | HS | 1.79 | –0.044 | –0.038 | 115% | –0.006 |

Generally, smaller absolute effects are expected for Ln systems as compared to TM ones, due to the semi-core nature of the f valence states. We note also that almost for each lanthanide system, the Ag-M interaction predominates the M-M one, being usually over an order of magnitude larger. Nevertheless, the introduction of Ag(II) in place of Cd(II) turned out to reveal some surprises. The effects for isotropic Gd(III) were rather small, and up to +23%. On the other hand, the Eu $f^6$ system with an $O^{2-}$ bridge experiences a change from FM to AFM Ln-Ln interaction due to $Ag^{2+}$ influence (in other words, the difference $J_{MM}$(Ag)–$J_{MM}$(Cd) is larger than the value of $J_{MM}$(Cd) itself). The largest influences in terms of absolute values are obtained for Ho and Yb clusters, reaching up to 0.21 $cm^{-1}$; for the former, which is equivalent to a 242% change relative to the Cd(II) reference cluster. Further enhancement can be anticipated to arise from proper engineering of the superexchange pathway and modifying the bridging ligands from fluoride to more spin-polarizable softer ligands.

A comparison of the *J* values for all molecules studied shows that in most cases the ground state corresponds to an AFM interaction between the M centers, consistent with a nearly linear M-X-M bridge geometry. However, the Ag-M interactions vary both qualitatively (AFM *vs*. FM) and quantitatively (i.e. from weak to strong) depending on the selected molecular system. This originates from differences in chemical bonding, geometry of the superexchange pathway for two lateral fluoride bridges (with the M-F-Ag angle not far from 100°, which represents the border between FM and AFM superexchange[44]), as well as a shape of spin density at M cations that depends on the auxiliary ligand field. Interestingly and counterintuitively, the M-M AFM interaction is sometimes enhanced in the presence of Ag(II) as compared to Cd(II) by a factor of up to 2.5. This is e.g. the case for M = Cu in clusters with X = F or O. In this particular case the spin density on central ligand switches its polarization from the p orbital, which is more or less parallel to the Cu-Cu vector, to the one which is close to a perpendicular one. Still, the fact that the Cu-X-Cu bridge is not perfectly linear permits spin interaction to take place rather than to cancel out. Thus, the influence of Ag(II) is in principle more complex than initially assumed, and may propagate via both a central and two auxiliary bridges.

The most interesting system seems to be the $Ni_2$(μ-O) cluster, where the Ni-Ni superexchange changes dramatically from initially strong AFM for Cd(II) (–100 $cm^{-1}$) to very strong FM for Ag(II) (+1670 $cm^{-1}$). The geometry and spin density in this complex (see SI) reveal that this system best exemplifies the success of the approach postulated here. First, the Ni-O and Ag-O bonds are very short, thus facilitating strong interactions. Secondly, Ag(II) introduces a huge spin density amounting to 0.49 $e^-$ onto the central oxo bridge, thus rendering it similar to a free radical $O^{\cdot-}$ (note that the transferred spin density in all other cases does not exceed 0.15 $e^-$). This obviously gives rise to a very strong FM superexchange, as expected, and qualitatively resembles an impact of organic free radical ligands. Quantitatively, the effect is dramatic, however, as the absolute value of $J_{MM}$ increases nearly 17 times in this case.

Overall, the impact of Ag(II) on the M–M superexchange is found to be moderate (up to 2.5 fold increase) and it enhances an antiferromagnetic character of $J_{MM}$ in nearly all cases. Aside from influencing the M–M superexchange, the presence of Ag introduces substantial Ag–M superexchange, which may be very strong and, thus, dictates the magnetic ground state of the corresponding molecule even, if spin frustration is present. However, in the case of $Ni_2$ oxo complex the effects are much larger and qualitatively opposite (FM instead of AFM superexchange). Comparison to the analogous $M_2$-Cd cases suggests that the key reason for the difference is in the fact that Ag(II) drastically polarizes the bridging ligands and introduces spin to the p orbitals, which are involved in the superexchange (**Figure 3**).

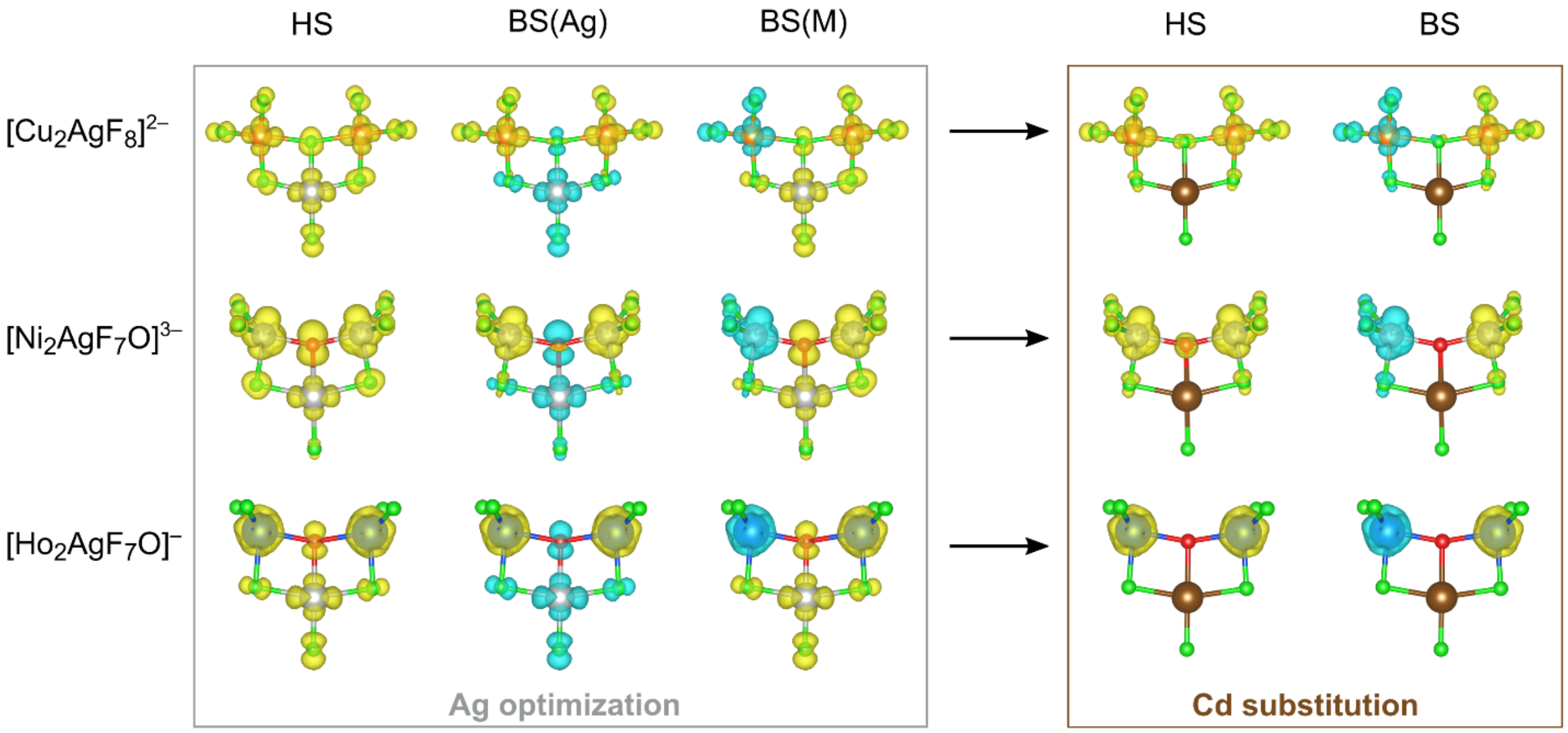

**Figure 3.** Spin densities for each spin state for three cluster geometries. Top row: Cu(II) with a $F^-$ bridge, center row: Ni(II) with an $O^{2-}$ bridge, bottom row: Ho(III) with an $O^{2-}$ bridge. The left panel shows spin densities for the optimized molecules (i.e. with Ag(II) cations), while after the effect of Ag(II) → Cd(II) substitution is shown in the right panel. Note the dramatic changes in the spin density of all bridging ligands associated with the presence of Ag(II) as compared to Cd(II) that greatly affect the M–M superexchange. The spin density isosurface value is set to 0.007 e/Å$^3$.

To get further insight into the importance of diverse superexchange pathways, we have looked into modified clusters with bridging ligands selectively removed.

Two main paths for M-M superexchange interaction exist in the model clusters: a short one through a central X ligand and a longer one via two side bridges and the Ag(II) cation. In order to clarify whether any of those is more important, we performed single-point calculations for a geometry corresponding to a previously optimized 'full' cluster, with either (i) the X ligand removed from the M-X-M path, or (ii) the two bridging fluorides eliminated from the M-F-Ag-F-M path. The abstraction is illustrated in **Figure 4** and the results are presented in **Table 3** (for geometry analysis cf. **Table S2** in **SI**).

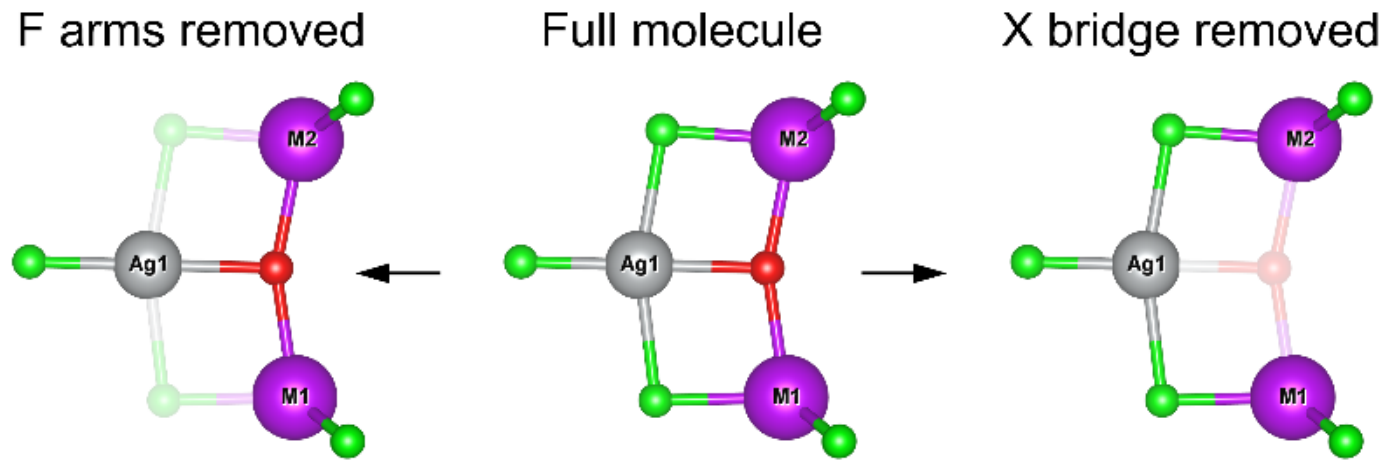

**Figure 4.** Scheme showing the abstraction of bridging ligands: two outer F atoms (left), or the central X atom (right). M stands for either 3d transition metal or lanthanide.

**Table 3.** Calculated exchange constants for full (optimized) molecules, and after abstraction of either the central X bridge, or the two outer F bridges from F-Ag-F pathways, in line with **Figure 4**. Data shows exchange constants provided in $cm^{-1}$, for M-M interaction before and after Ag → Cd substitution.

| | $[M_2AgF_7X]^{n-}$ (full molecule) | | | $[M_2AgF_7]^{n-}$ (X bridge removed) | | | $[M_2AgF_5X]^{n-}$ (outer F removed) | | |
|---|---|---|---|---|---|---|---|---|---|
| System | $J_{AgM}$ | $J_{MM}$(Ag) | $J_{MM}$(Cd) | $J_{AgM}$ | $J_{MM}$(Ag) | $J_{MM}$(Cd) | $J_{AgM}$ | $J_{MM}$(Ag) | $J_{MM}$(Cd) |
| Cu(F) | -285.5 | -127.6 | -84.7 | -602.3 | -29.4 | 4.1 | 428.3 | -236.6 | -8.1 |
| Gd(O) | -12.54 | -0.146 | -0.119 | 3.16 | 0.018 | 0.015 | -52.89 | -0.161 | -0.226 |
| Ni(O) | 409.1 | 1670.6 | -100.5 | 315.3 | -42.7 | -26.5 | 86.8 | 28.0 | 21.5 |

Let us focus on a Cu system with X = F. Here, for an initial cluster we have computed a ca. 51% increase (from *ca.* –85 to *ca.* –128 $cm^{-1}$) in the AFM Cu-Cu interaction upon the Cd(II) →Ag(II) substitution. It turns out that both types of ligand abstraction change the superexchange considerably: (i) in the absence of the central bridge the M-M superexchange is also antiferromagnetic, but much weaker than for initial system (ca. –29 $cm^{-1}$) since it is transferred via a longer pathway, while (ii) in the absence of the two outer fluoride ligands, the M-M superexchange is much stronger than for the initial cluster (ca. –237 $cm^{-1}$). Simultaneously, the corresponding Ag-M interactions are even more affected, since for (i) it more than doubles from *ca.* –286 to –602 $cm^{-1}$, while for (ii) it becomes strongly FM (ca. +428 $cm^{-1}$) as it now propagates via a bridge with a bond angle of nearly 90°. In any case, the effects of bridge removal are not additive (i.e. the complete system cannot be understood as a simple superposition of components lacking each type of bridging ligand at a time) and they certainly influence each other. The same conclusion holds for the other two clusters scrutinized (M = Gd with X = O and M = Ni with X = O, **Table 3**). Thus even for quite simple geometry and symmetry of the clusters, the complexity of their magnetic characteristics is substantial.

In view of noticeable or large effects seen in magnetic properties for Ag(II) systems, we pondered whether the Cu(II) cation could do the same job as Ag(II). This is of particular interest for Ln clusters, as well as for metal oxo clusters (as copper shows strong affinity to oxygen, as silver does to fluorine). Since dozens of structures of Cu(II)-Ln complexes have been described in the literature, we have also explored clusters containing Cu(II) for comparison. The cluster geometry after Ag(II) → Cu(II) substitution generally does not change much, and the superexchange coupling constants in the copper clusters could be calculated (**Table S5** in SI). It turns out that the $J_{AgM}$ values are usually 2 to 3 times larger than the corresponding $J_{CuM}$ ones. For example, Gd(III) oxide-bridged complex with Cu(II) exhibts $J_{AgM}$ of –3.64 $cm^{-1}$ while the Ag(II) shows $J_{AgM}$ of –12.54 $cm^{-1}$. Thus, the integration of Ag(II), rather than Cu(II), into Ln clusters, is more appealing in terms of the achieved increase in strength of magnetic interactions.

As this work is an attempt to motivate Ag(II), which can be seen as an inorganic analogue of an organic free radical, as a spin-polarizer, similar polynuclear spin clusterssystems incorporating organic radical bridges obviously constitute important reference systems. A number of lanthanide clusters with radical ligands were described in the literature over the past decade. One of the most astounding ones, described by Rinehart et al.[70], holds a $N_2^{\bullet 3-}$ ligand bridging two $Gd^{3+}$ centers, which are both coupled antiferromagnetically with the radical with a large exchange constant of –27 $cm^{-1}$. This value can be directly compared to our obtained $J_{AgM}$ values, which for the gadolinium cluster yields –12.5 $cm^{-1}$ with X = $O^{2-}$, which is of the same order of magnitude (whereas there is chemical bond between Gd and N in the published system, in our system Ag(II) is separated from Ln by nonmetal bridges). Unfortunately, the impact of the unusual free radical bridge, $N_2^{\bullet 3-}$, on the effective intercationic superexchange constant, $J_{MM}$, was not determined by the authors.[70] Yet another impressive attempt to maximize the radical-lanthanide exchange, ultimately resulted in a complex hosting antiferromagnetic coupling of –430 $cm^{-1}$ between $Yb^{3+}$ and a $bipy^{\bullet -}$ ligand ($2J$ = –920 $cm^{-1}$).[71] Unfortunately, this system contained only a single lanthanide center, so $J_{MM}$ is not relevant.

To compare the impact of Ag(II) and organic free radicals, we have run several calculations for systems containing a classical stable organic radical, TEMPO. The additional results presented in **Table S6** of SI show that the preferred lanthanide system is $Gd_2F_4O$ cluster with a TEMPO attached, where the O(2p) hole lies parallel to the Gd-O-Gd-O(TEMPO) plane. The TEMPO-$Gd^{3+}$ coupling can reach about 11.5 $cm^{-1}$, which is a comparable strength to $Ag^{2+}$-$Gd^{3+}$ coupling, although of an opposite, i.e. ferromagnetic character. On the other hand, the $[Ni_2F_5{\cdot}TEMPO]^{-}$ system shows remarkably strong TEMPO-Ni ferromagnetic coupling of +229 $cm^{-1}$, which is only about a half of the Ag(II)-Ni(II) coupling of 409 $cm^{-1}$. These results emphasize not only the powerful influence of $Ag^{2+}$ on the coupling between other cationic centers (i.e. via $J_{MM}$), but also for coupling itself as a $d^9$ radical to other transition metals (i.e. via $J_{AgM}$), and despite the fact that additional bridges separate Ag and the heterometal centers.

## CONCLUSIONS

Our calculations for small hypothetical coordination clusters containing two metal spin centers (M = TM or Ln) bridged by fluoride, oxo or chloro ligands, reveal that the presence of a divalent silver cation can substantially alter the cluster's magnetic characteristics. Typically, Ag(II) induces a strong polarization to any ligand attached. If tightly connected to the bridging ligand that mediates the M–M superexchange, as is the cae in our model clusters, Ag(II) may enhance the M–M superexchange resulting in exchange constants that increase in absolute terms up to 0.2 $cm^{-1}$ for lanthanides (i.e. often by a factor of two), and even more than 200 $cm^{-1}$ for 3d transition metals. In the extreme case of a $Ni_2$ complex with a central oxo bridge, the spin-polarizing effect is so large that it renders the oxo bridge similar to a radical anion and can even change the nature of the interaction (from AFM to FM). Simultaneously, the Ag(II)–M superexchange is typically very strong and occasionally leads to spin frustration in the some of the studied clusters. For example, the Ag(II)–Ln superexchange may be comparable to, or even surpass the Cu(II)-Ln exchange reported in the literature, reaching $J$ values as high as –31.4 $cm^{-1}$, efficiently increasing the gaps between energy levels of the Ln elements. These results indicate a potential path toward more complex polynuclear clusters with exchange-coupled spin centers that in principle can result in significantly higher multiplet states. Future work will focus on deconvoluting the various coupling mechanisms further and comparing to an extended set of structural analogues. Ag(II) seems to offer an exciting alternative to organic radical ligands in this field.

## ASSOCIATED CONTENT

**Supporting Information**. Comprehensive computational details, validation of methodology, molecular geometries including superexchange pathways, Löwdin spin populations and analysis of superexchange interactions, including those for selected Cu(II) reference systems, may be found in the Supporting Information (**SI**).

AUTHOR INFORMATION

### Corresponding Author

* Wojciech Grochala, Center of New Technologies, University of Warsaw, Zwirki i Wigury 93, 02089 Warsaw Poland, e-mail: w.grochala@cent.uw.edu.pl

* Paul Kögerler, Institut für Anorganische Chemie, RWTH Aachen University, 52056 Aachen, Germany, e-mail: paul.koegerler@ac.rwth-aachen.de

### Author Contributions

The manuscript was written through contributions of all authors. All authors have given approval to the final version of the manuscript.

## ACKNOWLEDGMENT

WG and PK jointly acknowledge Polish National Science Center (NCN) and Deutsche Forschungsgemeinschaft (DFG) for funding under the collaborative project Beethoven (2016/23/G/ST5/04320 or KO 3990/8-1). This research was supported by the Interdisciplinary Centre for Mathematical and Computational Modelling (ICM), University of Warsaw (grants GA83-34 and G49-17).

This work is dedicated to P. Pyykkö on his 80th birthday.

TOC Graphic

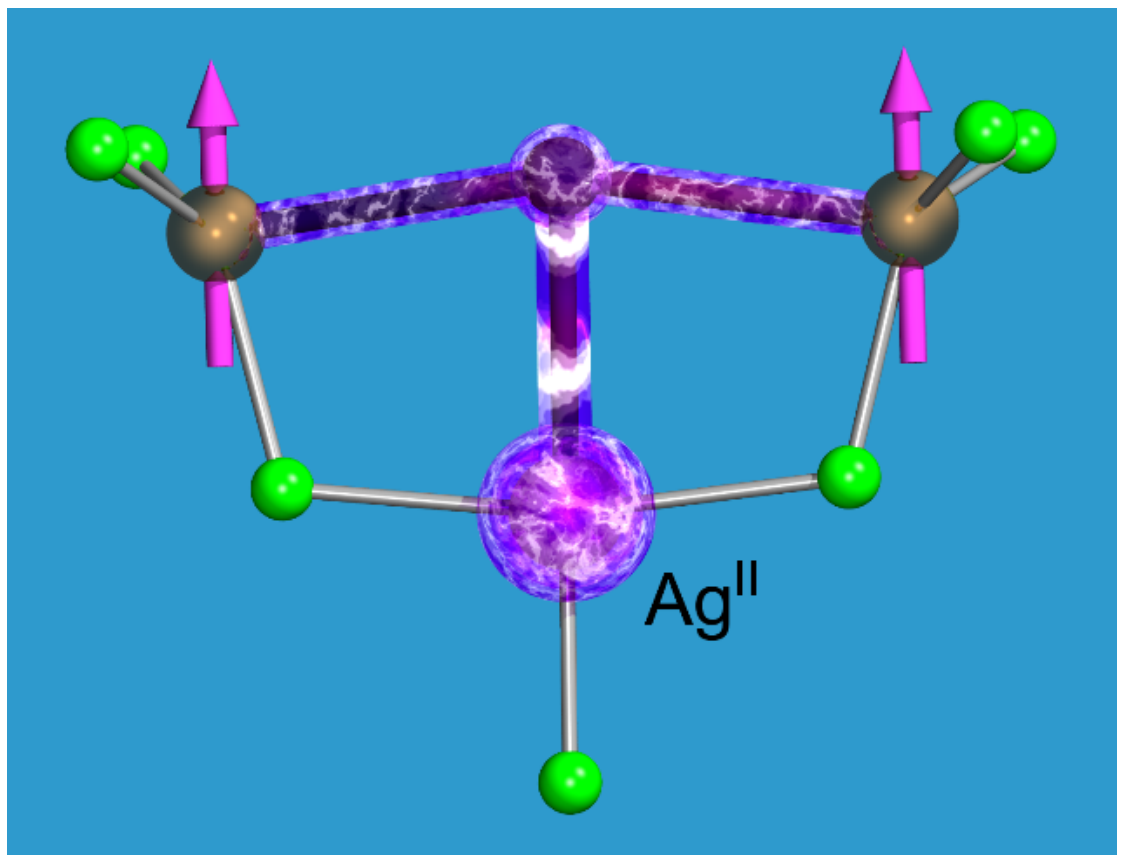

TOC description:

" Divalent silver is capable to spin-polarize monoatomic ligands bridging two transition metal or lanthanide cations. By inducing free radical character, it may immensely influence the magnetic superexchange coupling between these metal centers."

# Ag(II) as Spin Super-Polarizer in Molecular Spin Clusters

Mateusz Domański[a], Jan van Leusen[b], Marvin Metzelaars,[b] Paul Kögerler*[b], and Wojciech Grochala*[a]

[a] Prof. Wojciech Grochala, Mateusz Domański
Center of New Technologies
University of Warsaw
Zwirki i Wigury 93, 02089 Warsaw Poland
E-mail: w.grochala@cent.uw.edu.pl

[b] Dr. Marvin Matzelaars, Dr. Jan van Leusen, Prof. Paul Kögerler
Institut für Anorganische Chemie
RWTH Aachen University,
52056 Aachen, Germany
E-mail: paul.koegerler@ac.rwth-aachen.de

## 1. Methods

The largest effort in theoretical research on magnetic clusters containing lanthanides is spent for a correct description of 4f electrons. These levels are subject to a splitting originating from ligand (crystal) field and spin-orbit interactions.[1] Recently there have been published many computational studies on spin-spin coupling in molecular systems containing one or more lanthanide cations, i.a. : cerium[2], praseodymium[3], europium[4], gadolinium[4], terbium[4,5], dysprosium[4,6], holmium[4,6]. In all of these studies[2–6] the spin-orbit coupling effects have been considered. However, for the most gadolinium studies[7–14] where the spin-orbit coupling effect is predicted to be weak, the scalar-relativistic methods and the Ising type Hamiltonian were used. The issue with spin-orbit coupling is that it can be relatively easy employed for single centre systems, and then usually CAS is nowadays used for this. However, on larger systems this approach becomes often too expensive, but still there is a necessity to correctly account for relativistic effects[9]. Due to extensive range of hypothetical molecules considered in this study, we chose an approach of robust hybrid DFT approach together with a scalar relativistic Hamiltonian and all-electron basis set.

The study is based on density functional theory (DFT) calculations using ORCA 4.2.1. program[15] applying the frequently used B3LYP functional[16,17]. Calculations were performed with a "very tight" SCF convergence procedure and with the most accurate numerical integration grid available (Lebedev 770 points). The relativistic effects were included with the use of $0^{th}$ order regular approximation (ZORA) Hamiltonian[18,19] with one-centre approximation as implemented in ORCA. The ZORA Hamiltonian including spin-orbit effects is only available for closed-shell molecules in DFT calculations, thus we applied scalar relativistic ZORA Hamiltonian. Segmented all-electron relativistically contracted basis set SARC-ZORA-TZVP was used for Ag and lanthanides and relativistically recontracted ZORA-def2-TZVP basis for the lighter elements[20,21]. Moreover, properties calculated with the SARC-ZORA-TZVP compact basis set are consistent with experimentally measured properties for heavy elements and their molecules, *i.a.* ionization potentials, despite the calculations were performed with scalar relativistic Hamiltonian ZORA[20,21]. Especially, in contrast to heavier elements, the effects of spin-orbit coupling in the lanthanides are not usually considered of critical importance because they are smaller in magnitude than the correlation effects arising from the different electronic configurations.[21–24] Moreover, for the methods we use here (hybrid DFT B3LYP), it is not expected that inclusion of spin-orbit corrections would lead to systematic improvement of the computed values. [21–24] Thus, the spin-orbit coupling was neglected to focus on the impact of $Ag^{2+}$ on the 4f cations interactions.

To calculate the strength of spin-spin interactions the Ising model Hamiltonian was used in the form $H = -J_{\langle ij\rangle} \sum_{i,j} S_{z,i} S_{z,j}$ (with $J_{\langle ij\rangle}$=$J_{\langle ji\rangle}$), where $J_{\langle ij\rangle}$ is antiferromagnetic coupling constant between the closest neighbouring magnetic centres $i$ and $j$ when $J_{ij} < 0$ . The exchange spin coupling parameters $J$ were calculated using the "broken symmetry" (BS) formalism[25] by calculation of all the possible spin states in each system and solving the set of linear equations. Due to the fact that only a part of molecular geometries were perfectly symmetric (**Table S2**), the exchange constants between silver and two other cations in principle are not equal. Because of that, in the model molecules of $AgM_2F_7X$ (X = F, Cl, O) three types of exchange constants were applied, $J_{Ag\text{-}M1}$ , $J_{Ag\text{-}M2}$ and $J_{M\text{-}M}$, these results are presented in **Table S3**.

For each chemical system, the geometry optimization for each possible spin-state was performed in a search for the ionic and electronic ground state simultaneously. To apply broken symmetry formalism, the energies of excited spin states were calculated in a ground state molecular geometry. For every system, possible ground states are; high spin HS, silver broken-symmetry state BS(Ag) (with Ag having opposite spin) and the other metal broken-symmetry state BS(M) (with M1 or M2 having opposite spin). Importantly, for each spin state the energy was thoroughly optimized to the lowest value, in particular by ensuring that all of the orbitals are occupied in the same manner. This consistency further diminishes effect of spin-orbit coupling, as all spin-states would be affected with the same way since all orbitals are similarly occupied both in Ag and in Cd-substituted complexes. To check orbital occupancy, Löwdin spin population analyses[26] were carried out (as implemented in Orca), with f-orbital spin populations carefully inspected during calculations, see **Table S7**, **Table S8**, **Table S9**, **Table S10**, **Table S11** and **Table S12**. Finally, to evaluate the influence of $Ag^{2+}$ spin polarizing properties on the f-electrons, the $Ag^{2+}$ cations were substituted with $Cd^{2+}$ cations. In each substituted system $J_{M\text{-}M}$ was recalculated – changes upon substitution indicate, whether or not, $Ag^{2+}$ influences this interaction (**Table 1** in the main paper). To highlight influence of Ag, we also presented spin population on the key atoms, see **Table S4**.

Finally, we have conducted a series of calculations to validate our methodology. Since similar approaches have been successfully utilized in the past for 3d and 4d spin interactions, we assume that in these cases the methods are robust. Thus, here the focus has been focused especially on Ln-Ln and TM-Ln interactions. We applied the method described above for the molecules studied previously in the literature for which experimental data is known. **Table S1** shows comparison of our results to the reference data. The B3LYP/ZORA calculated parameters show good qualitative and fair semi-quantitative agreement with experimental data, the latter itself often having a considerable margin of error (though we notice that experimental errors are rarely provided). Computations with the chosen method yields results within ~1 $cm^{-1}$ despite the fact, that molecular geometries of the complexes were not optimized here, and periodic character of the crystalline systems was ignored. Since the current study relies on Ag/Cd and Cu/Zn substitutions, substantial error cancelling is expected and the major property trends are clearly detectable using our approach.

**Table S1.** Comparison of results calculated here (B3LYP/ZORA) to reference data of TM-Ln or Ln-Ln complexes.

| Molecule | Basis | $J(Cu\text{-}Ln)^{calc}$ / $cm^{-1}$ | $J(Cu\text{-}Ln)^{ref}$ / $cm^{-1}$ | $J(Ln\text{-}Ln)^{calc}$ / $cm^{-1}$ | $J(Ln\text{-}Ln)^{ref}$ / $cm^{-1}$ | Δ / $cm^{-1}$ | Ref. |
|---|---|---|---|---|---|---|---|
| Gd Cu $C_{12}$ $H_{16}$ $N_5$ $O_{13}$ | SARC-TZVP(Gd), TZVP(Cu, C, H, N, O) | 4.79 | 3.36 [a] | - | - | 1.43 | [27] |
| Gd Cu $C_{45}$ $H_{65}$ $N_2$ $O_{11}$ | SARC-TZVP(Gd), TZVP(Cu,N,O), SVP(C,H) | 3.20 | 4.20 [b] | - | - | -1.00 | [28] |
| Gd Cu $C_{32}$ $H_{38}$ $F_9$ $N_2$ $O_{12}$ | SARC-TZVP(Gd), TZVP(Cu, C, H, F, N, O) | 5.14 | 4.42 [b] | - | - | 0.72 | [29] |
| Gd $Cu_2$ $C_{40}$ $H_{32}$ $F_9$ $N_4$ $O_{10}$ | SARC-TZVP(Gd), TZVP(Cu, C, H, F, N, O) | 2.75 | 3.54 [b] | - | - | -0.79 | [29] |
| $Tb_2$ $C_{40}$ $H_{46}$ $N_8$ $O_{14}$ | SARC-TZVP(Tb), TZVP(N,O), SVP(C, H) | - | - | -0.43 | -0.10 [c] | -0.33 | [4] |
| $Gd_2$ $C_{40}$ $H_{46}$ $N_8$ $O_{14}$ | SARC-TZVP(Gd), TZVP(N,O), SVP(C, H) | - | - | -0.28 | -0.08 [c] | -0.20 | [4] |

[a] experimental value, based on magnetostructural correlation of dihedral angle and *J*, correl. coeff. $R^2$ = 0.9145.

[b] experimental value, least-square fitting to magnetization data, no correlation coefficient provided.

[c] value calculated on CASSCF/DKH level of theory with Lines Hamiltonian including Spin-Orbit coupling, which results fit the experimental data in the reference[4]. No correlation coefficients provided.

## 2. Additional data

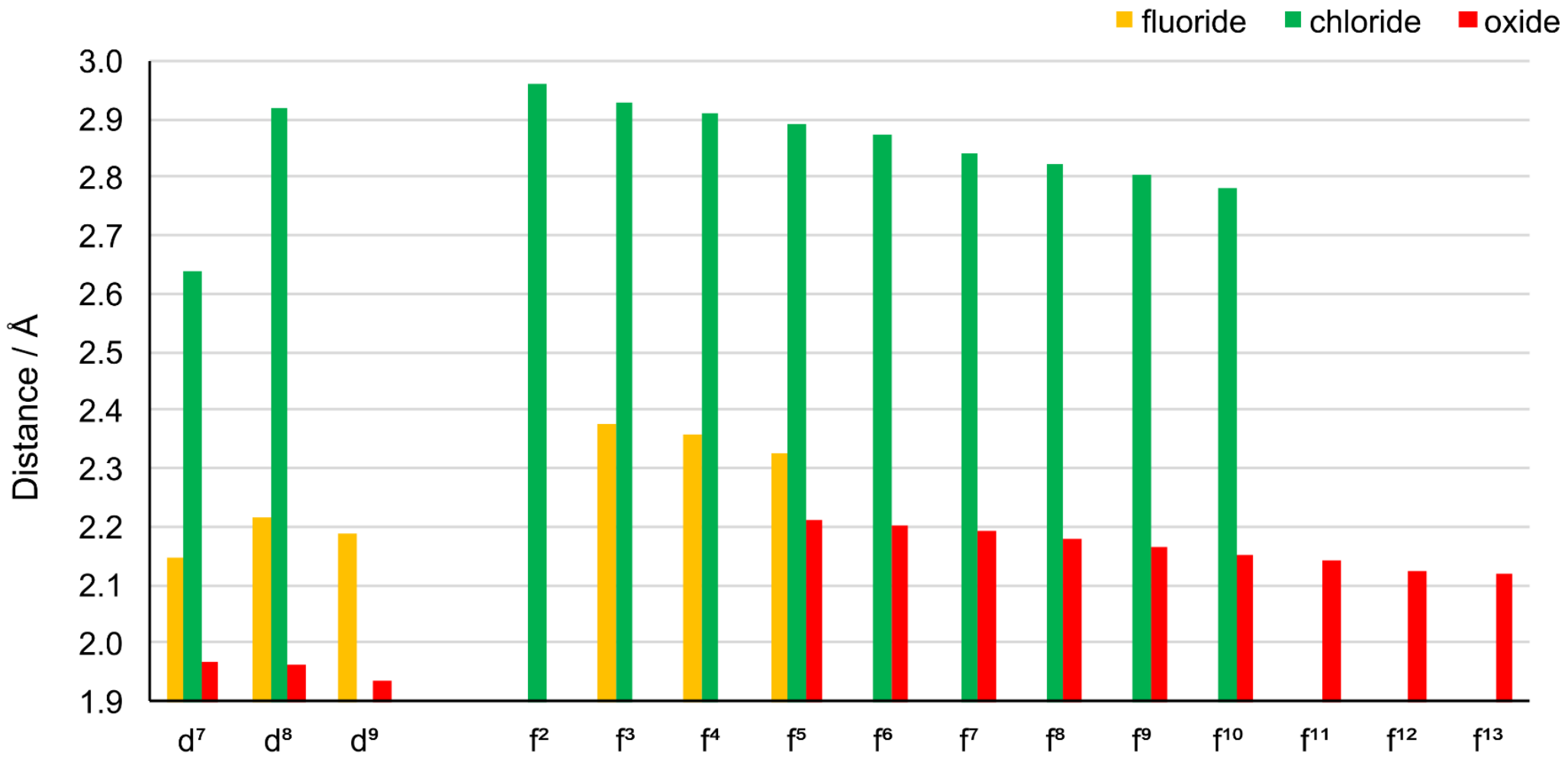

**Figure S1.** Lengths of Ag-X bonds (X – a bridging ligand) in the optimized clusters studied in this work.

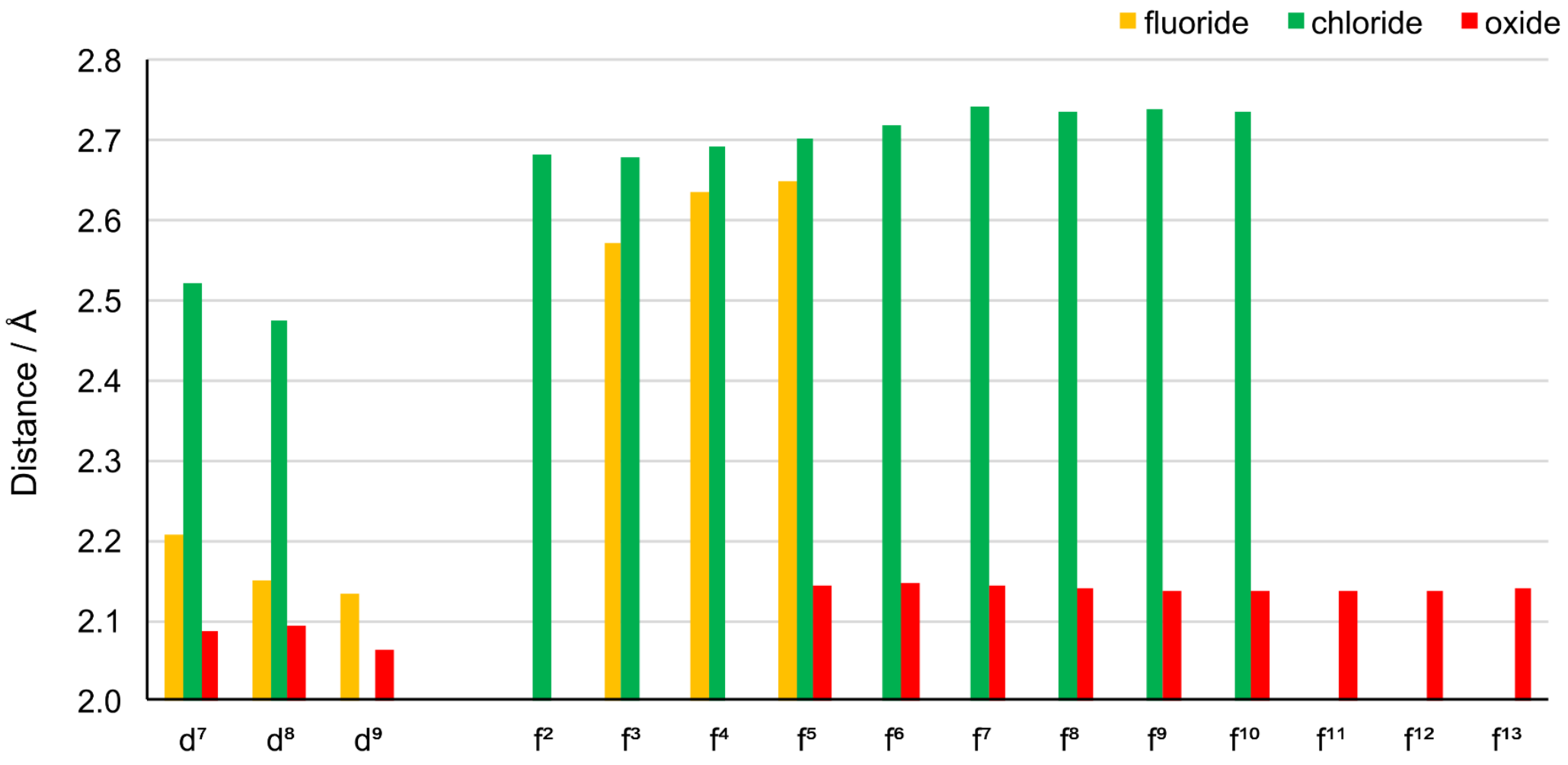

**Figure S2.** Average length of M-X bonds (X – a bridging ligand) in the optimized clusters studied in this work.

**Table S2.** The key bond distances (in Å) an angles (in degrees) in the optimized clusters studied in this work (X – a bridging ligand). Difference d($M^1$-x)–d($M^2$-X) shows the asymmetry of the clusters. Symbols in square brackets indicate the next element in the path.

| | | | | Ag-X-$M^1$, Ag-X-$M^2$ and $M^1$-X-$M^2$ bridges | | | | | | $M^1$-F-Ag bridge | | | $M^2$-F-Ag bridge | | | | | |
|---|---|---|---|---|---|---|---|---|---|---|---|---|---|---|---|---|---|---|
| Ligand | M | v.e. | GS | d(Ag-X) | d($M^1$-X) | d($M^2$-X) | ∠ ($M^2$-X-$M^1$) | ∠ ($M^1$-X-Ag) | ∠ ($M^2$-X-Ag) | d($M^1$-F) [-Ag] | d(Ag-F) [-$M^1$] | ∠ ($M^1$-F-Ag) | d($M^2$-F) [-Ag] | d(Ag-F) [-$M^2$] | ∠ ($M^2$-F-Ag) | d($M^1$-Ag) | d($M^2$-Ag) | Difference d($M^1$-X)-d($M^2$-X) |
| | Co | $d^7$ | BS(Ag) | 2.206 | 2.146 | 2.147 | 168.6 | 96.4 | 94.9 | 2.032 | 2.131 | 102.4 | 2.014 | 2.136 | 101.2 | 3.246 | 3.208 | -0.001 |
| | Ni | $d^8$ | BS(Ag) | 2.149 | 2.215 | 2.216 | 168.4 | 95.8 | 95.8 | 1.984 | 2.131 | 103.4 | 1.985 | 2.131 | 103.5 | 3.238 | 3.240 | -0.001 |
| $F^-$ | Cu | $d^9$ | BS(Ag) | 2.132 | 2.188 | 2.186 | 164.2 | 97.8 | 98.0 | 1.987 | 2.131 | 104.5 | 1.986 | 2.131 | 104.6 | 3.257 | 3.259 | 0.002 |
| | Nd | $f^3$ | BS(Ln) | 2.571 | 2.378 | 2.375 | 168.9 | 95.4 | 95.7 | 2.263 | 2.143 | 112.4 | 2.268 | 2.143 | 112.6 | 3.662 | 3.671 | 0.003 |
| | Pm | $f^4$ | BS(Ag) | 2.635 | 2.355 | 2.356 | 171.9 | 94.3 | 94.3 | 2.249 | 2.142 | 113.1 | 2.250 | 2.141 | 113.1 | 3.664 | 3.664 | -0.001 |
| | Sm | $f^5$ | HS | 2.647 | 2.325 | 2.325 | 171.2 | 94.4 | 94.4 | 2.238 | 2.144 | 113.1 | 2.237 | 2.143 | 113.1 | 3.655 | 3.655 | 0.000 |
| | Co | $d^7$ | BS(Ag)* | 2.086 | 1.969 | 1.969 | 156.2 | 101.9 | 101.9 | 2.086 | 2.275 | 92.4 | 2.086 | 2.275 | 92.4 | 3.150 | 3.150 | 0.000 |
| | Ni | $d^8$ | HS | 2.093 | 1.962 | 1.962 | 154.9 | 102.5 | 102.5 | 2.027 | 2.296 | 93.9 | 2.027 | 2.296 | 93.9 | 3.164 | 3.164 | 0.000 |
| | Cu | $d^9$ | BS(Ag)* | 2.064 | 1.934 | 1.934 | 149.2 | 105.4 | 105.4 | 2.111 | 2.224 | 94.3 | 2.099 | 2.227 | 94.7 | 1.181 | 3.180 | 0.000 |
| | Sm | $f^5$ | BS(Ag) | 2.142 | 2.209 | 2.209 | 159.4 | 100.3 | 100.3 | 2.210 | 2.281 | 96.1 | 2.210 | 2.281 | 96.1 | 3.341 | 3.340 | 0.000 |
| | Eu | $f^6$ | BS(Ag) | 2.147 | 2.201 | 2.201 | 160.9 | 99.6 | 99.5 | 2.203 | 2.284 | 95.5 | 2.203 | 2.283 | 95.5 | 3.320 | 3.320 | 0.000 |
| $O^{2-}$ | Gd | $f^7$ | BS(Ag) | 2.142 | 2.192 | 2.192 | 160.5 | 99.8 | 99.8 | 2.186 | 2.289 | 95.5 | 2.186 | 2.289 | 95.6 | 3.314 | 3.314 | 0.000 |
| | Tb | $f^8$ | BS(Ln) | 2.140 | 2.181 | 2.173 | 160.2 | 99.9 | 99.9 | 2.164 | 2.293 | 95.8 | 2.175 | 2.290 | 95.3 | 3.308 | 3.301 | 0.008 |
| | Dy | $f^9$ | BS(Ln) | 2.138 | 2.166 | 2.165 | 160.0 | 99.9 | 100.1 | 2.156 | 2.290 | 95.6 | 2.155 | 2.290 | 95.8 | 3.295 | 3.299 | 0.001 |
| | Ho | $f^{10}$ | HS | 2.137 | 2.152 | 2.152 | 160.2 | 99.9 | 99.9 | 2.144 | 2.292 | 95.4 | 2.144 | 2.291 | 95.4 | 3.283 | 3.282 | 0.000 |
| | Er | $f^{11}$ | BS(Ag) | 2.137 | 2.142 | 2.142 | 160.7 | 99.7 | 99.7 | 2.129 | 2.293 | 95.3 | 2.129 | 2.293 | 95.3 | 3.269 | 3.270 | 0.000 |
| | Tm | $f^{12}$ | HS | 2.136 | 2.125 | 2.125 | 160.7 | 99.7 | 99.6 | 2.115 | 2.298 | 95.0 | 2.114 | 2.299 | 95.0 | 3.256 | 3.255 | 0.000 |
| | Yb | $f^{13}$ | HS | 2.139 | 2.118 | 2.119 | 161.7 | 99.2 | 99.2 | 2.105 | 2.301 | 94.6 | 2.105 | 2.301 | 94.6 | 3.241 | 3.241 | -0.001 |
| | Co | $d^7$ | BS(M) | 2.520 | 2.643 | 2.637 | 163.9 | 82.2 | 81.8 | 1.990 | 2.147 | 110.2 | 1.984 | 2.150 | 109.4 | 3.394 | 3.376 | 0.006 |
| | Ni | $d^8$ | BS(Ag) | 2.473 | 2.918 | 2.917 | 159.7 | 79.9 | 79.9 | 1.935 | 2.167 | 115.9 | 1.935 | 2.166 | 115.8 | 3.478 | 3.476 | 0.001 |
| | Pr | $f^2$ | BS(Ag) | 2.683 | 2.961 | 2.960 | 169.1 | 84.5 | 84.6 | 2.243 | 2.180 | 118.5 | 2.244 | 2.181 | 118.5 | 3.802 | 3.802 | 0.001 |
| | Nd | $f^3$ | BS(Ln) | 2.680 | 2.929 | 2.925 | 169.4 | 84.6 | 84.8 | 2.228 | 2.178 | 118.0 | 2.233 | 2.179 | 118.1 | 3.778 | 3.785 | 0.004 |
| | Pm | $f^4$ | BS(Ag) | 2.691 | 2.908 | 2.908 | 168.5 | 84.3 | 84.3 | 2.214 | 2.178 | 117.7 | 2.214 | 2.178 | 117.7 | 3.759 | 3.759 | 0.000 |
| $Cl^-$ | Sm | $f^5$ | BS(Ln) | 2.702 | 2.902 | 2.877 | 168.1 | 83.8 | 84.3 | 2.203 | 2.176 | 117.6 | 2.200 | 2.180 | 117.5 | 3.745 | 3.744 | 0.025 |
| | Eu | $f^6$ | HS | 2.719 | 2.873 | 2.873 | 167.2 | 83.6 | 83.6 | 2.186 | 2.176 | 117.5 | 2.186 | 2.176 | 117.4 | 3.728 | 3.728 | 0.000 |
| | Gd | $f^7$ | HS | 2.741 | 2.840 | 2.840 | 167.1 | 83.6 | 83.6 | 2.175 | 2.174 | 117.6 | 2.175 | 2.174 | 117.6 | 3.720 | 3.720 | 0.000 |
| | Tb | $f^8$ | HS | 2.734 | 2.830 | 2.817 | 167.1 | 83.4 | 83.7 | 2.155 | 2.178 | 117.3 | 2.165 | 2.178 | 117.1 | 3.701 | 3.705 | 0.013 |
| | Dy | $f^9$ | HS | 2.738 | 2.804 | 2.804 | 167.0 | 83.5 | 83.5 | 2.146 | 2.178 | 117.2 | 2.146 | 2.178 | 117.2 | 3.691 | 3.691 | 0.000 |
| | Ho | $f^{10}$ | HS | 2.734 | 2.781 | 2.781 | 167.0 | 83.5 | 83.5 | 2.134 | 2.181 | 116.7 | 2.134 | 2.180 | 116.7 | 3.673 | 3.673 | 0.000 |

**Table S3.** Computed data for each considered system: valence electron configuration (v.e.), ground state configuration (GS), values of silver-metal and metal-metal superexchange constants, ($J_{MM}$ also after Ag(II) → Cd(II) substitution). Silver-metal exchange couplings are here separated to account for asymmetry in molecular geometries.

| Ligand | Metal | v.e. | GS | $J_{AgM(1)}$ / cm$^{-1}$ | $J_{AgM(2)}$ / cm$^{-1}$ | $J_{AgM(1)} - J_{AgM(2)}$ / cm$^{-1}$ | $J_{MM}$(Ag) / cm$^{-1}$ | $J_{MM}$(Cd) / cm$^{-1}$ |
|---|---|---|---|---|---|---|---|---|
| $F^-$ | Co | $d^7$ | BS(Ag) | 23.106 | 58.964 | -35.86 | -17.252 | -18.161 |
| | Ni | $d^8$ | BS(Ag) | -43.247 | -45.562 | 2.31 | -12.482 | -11.128 |
| | Cu | $d^9$ | BS(Ag) | -278.009 | -292.910 | 14.90 | -127.620 | -84.671 |
| | Nd | $f^3$ | BS(Ln) | -19.678 | 0.702 | -20.38 | -0.112 | -0.175 |
| | Pm | $f^4$ | BS(Ag) | -1.752 | -1.801 | 0.05 | -0.278 | -0.262 |
| | Sm | $f^5$ | HS | 2.686 | 2.696 | -0.01 | -0.185 | -0.194 |
| $O^{2-}$ | Co | $d^7$ | BS(Ag)* | 326.472 | 326.450 | 0.02 | -71.121 | -31.086 |
| | Ni | $d^8$ | HS | 409.157 | 408.949 | 0.21 | 1670.595 | -100.520 |
| | Cu | $d^9$ | BS(Ag)* | -215.527 | -294.961 | 79.43 | -608.671 | -380.953 |
| | Sm | $f^5$ | BS(Ag) | -30.147 | -29.743 | -0.40 | -1.073 | -1.018 |
| | Eu | $f^6$ | BS(Ag) | -22.619 | -22.588 | -0.03 | -0.016 | 0.084 |
| | Gd | $f^7$ | BS(Ag) | -12.555 | -12.531 | -0.02 | -0.146 | -0.119 |
| | Tb | $f^8$ | BS(Ln) | 2.465 | -11.068 | 13.53 | -0.160 | -0.130 |
| | Dy | $f^9$ | BS(Ln) | -1.691 | -1.239 | -0.45 | -0.293 | -0.197 |
| | Ho | $f^{10}$ | HS | 2.867 | 2.827 | 0.04 | -0.281 | -0.116 |
| | Er | $f^{11}$ | BS(Ag) | -9.050 | -8.936 | -0.11 | -0.289 | -0.131 |
| | Tm | $f^{12}$ | HS | 21.777 | 21.633 | 0.14 | -0.893 | -0.826 |
| | Yb | $f^{13}$ | HS | 19.683 | 19.537 | 0.15 | 0.326 | 0.540 |
| $Cl^-$ | Co | $d^7$ | BS(M) | 3.495 | 11.856 | -8.36 | -32.337 | -31.521 |
| | Ni | $d^8$ | BS(Ag) | -83.176 | -82.998 | -0.18 | -2.653 | -1.136 |
| | Pr | $f^2$ | BS(Ag) | -31.398 | -31.477 | 0.08 | -0.395 | -0.334 |
| | Nd | $f^3$ | BS(Ln) | -12.481 | 3.531 | -16.01 | -0.361 | -0.196 |
| | Pm | $f^4$ | BS(Ag) | -1.943 | -1.953 | 0.01 | -0.279 | -0.275 |
| | Sm | $f^5$ | BS(Ln) | 1.530 | 2.680 | -1.15 | -0.090 | -0.122 |
| | Eu | $f^6$ | HS | 2.893 | 2.923 | -0.03 | -0.087 | -0.080 |
| | Gd | $f^7$ | HS | 2.106 | 2.111 | 0.00 | -0.096 | -0.092 |
| | Tb | $f^8$ | HS | 2.070 | 7.780 | -5.71 | -0.095 | -0.090 |
| | Dy | $f^9$ | HS | 6.880 | 6.880 | 0.00 | -0.078 | -0.071 |
| | Ho | $f^{10}$ | HS | 1.662 | 1.917 | -0.26 | -0.044 | -0.038 |

**Table S4.** Löwdin spin populations projected on the key atoms in [$AgM_2F_7X$] clusters in a high-spin state. Spin populations for Cd-substituted molecules (only of high-spin state) together with the ratio highlight the impact of $Ag^{2+}$ on bridging atom spin-polarization. In case of cluster asymmetry, the M spin populations were averaged.

| Ligand | Metal | v.e. | GS | Ag | M | $F^{lone}$ | $X^{bridge}$(Ag) | $X^{bridge}$(Cd) | $X^{bridge}$(Ag) /$X^{bridge}$(Cd) |
|---|---|---|---|---|---|---|---|---|---|
| $F^-$ | Co | $d^7$ | BS(Ag) | 0.516 | 2.791 | 0.190 | 0.096 | 0.048 | 200% |
| | Ni | $d^8$ | BS(Ag) | 0.551 | 1.792 | 0.183 | 0.094 | 0.020 | 470% |
| | Cu | $d^9$ | BS(Ag) | 0.545 | 0.742 | 0.172 | 0.119 | 0.037 | 322% |
| | Nd | $f^3$ | BS(Ln) | 0.487 | 3.024 | 0.391 | 0.016 | -0.004 | -400% |
| | Pm | $f^4$ | BS(Ag) | 0.482 | 4.021 | 0.402 | 0.013 | -0.004 | -325% |
| | Sm | $f^5$ | HS | 0.482 | 5.033 | 0.404 | 0.009 | -0.007 | -129% |
| $O^{2-}$ | Co | $d^7$ | BS(Ag)* | 0.382 | 2.818 | 0.024 | 0.484 | 0.140 | 346% |
| | Ni | $d^8$ | HS | 0.391 | 1.817 | 0.024 | 0.491 | 0.114 | 431% |
| | Cu | $d^9$ | BS(Ag)* | 0.437 | 0.757 | 0.063 | 0.124 | 0.037 | 335% |
| | Sm | $f^5$ | BS(Ag) | 0.522 | 5.039 | 0.170 | 0.106 | -0.030 | -353% |
| | Eu | $f^6$ | BS(Ag) | 0.519 | 6.051 | 0.172 | 0.104 | -0.030 | -347% |
| | Gd | $f^7$ | BS(Ag) | 0.524 | 7.023 | 0.171 | 0.119 | -0.017 | -700% |
| | Tb | $f^8$ | BS(Ln) | 0.519 | 6.006 | 0.167 | 0.140 | -0.005 | -2800% |
| | Dy | $f^9$ | BS(Ln) | 0.523 | 4.993 | 0.167 | 0.142 | 0.003 | 4733% |
| | Ho | $f^{10}$ | HS | 0.522 | 3.999 | 0.167 | 0.143 | 0.002 | 7150% |
| | Er | $f^{11}$ | BS(Ag) | 0.524 | 3.004 | 0.168 | 0.137 | -0.001 | -13700% |
| | Tm | $f^{12}$ | HS | 0.524 | 3.004 | 0.168 | 0.137 | 0.006 | 2283% |
| | Yb | $f^{13}$ | HS | 0.523 | 0.992 | 0.169 | 0.142 | 0.004 | 3550% |
| $Cl^-$ | Co | $d^7$ | BS(M) | 0.498 | 2.773 | 0.190 | 0.138 | 0.042 | 329% |
| | Ni | $d^8$ | BS(Ag) | 0.551 | 1.792 | 0.183 | 0.094 | 0.020 | 470% |
| | Pr | $f^2$ | BS(Ag) | 0.482 | 2.033 | 0.346 | 0.044 | -0.005 | -880% |
| | Nd | $f^3$ | BS(Ln) | 0.483 | 3.028 | 0.345 | 0.044 | -0.005 | -880% |
| | Pm | $f^4$ | BS(Ag) | 0.480 | 4.023 | 0.351 | 0.042 | -0.006 | -700% |
| | Sm | $f^5$ | BS(Ln) | 0.479 | 5.041 | 0.354 | 0.040 | -0.007 | -571% |
| | Eu | $f^6$ | HS | 0.477 | 6.056 | 0.359 | 0.040 | -0.006 | -667% |
| | Gd | $f^7$ | HS | 0.477 | 7.016 | 0.365 | 0.037 | -0.007 | -529% |
| | Tb | $f^8$ | HS | 0.476 | 5.995 | 0.365 | 0.039 | -0.005 | -780% |
| | Dy | $f^9$ | HS | 0.476 | 4.984 | 0.367 | 0.041 | -0.003 | -1367% |
| | Ho | $f^{10}$ | HS | 0.477 | 3.990 | 0.368 | 0.041 | -0.002 | -2050% |

**Table S5.** Computed data for several Cu analogues **(a)** of Ag systems **(b)**: valence electron configuration (v.e.), ground state configuration (GS), values of silver-metal and metal-metal superexchange constants before and after Cu(II)→Zn(II) substitution ($J_{CuM}$, $J_{MM}$(Cu) and $J_{MM}$(Zn), respectively) or correspondingly Ag(II)→Cd(II) substitution. Each system has one GS out of three possible spin-states: high-spin "HS", broken-symmetry with S(Cu) of opposite polarity "BS(Cu)", or broken-symmetry with S(M) of opposite polarity "BS(M)", correspondingly for Ag. Ratio is defined as Ratio=$J_{MM}$(Cu)/$J_{MM}$(Zn), while difference as Δ=$J_{MM}$(Cu)–$J_{MM}$(Zn), or correspondingly with Ag and Cd. Clusters in principle may lack symmetry elements thus $J_{CuM}$ may be an average over two interactions (Cu–$M^1$ and Cu–$M^2$). Each copper cluster was optimized in a ground state electronic configuration, with 3d and 4f electrons' configurations in the excited states checked to be equivalent to GS **(c)**.

**a)** Copper clusters $[M_2CuF_8]^{2-}$ (M = Ni, Co), $[M_2CuF_7O]^{3-}$ (M = Ni, Co) and $[M_2CuF_7O]^{-}$ (M = Eu, Gd, Tb)

| Ligand | Metal | v.e. | GS | $J_{CuM}$ / cm$^{-1}$ | $J_{MM}$(Cu) / cm$^{-1}$ | $J_{MM}$(Zn) / cm$^{-1}$ | Ratio | Δ / cm$^{-1}$ |
|---|---|---|---|---|---|---|---|---|
| $F^-$ | Co | $d^7$ | BS(M) | –12.4 | –15.4 | –15.4 | 100% | 0.0 |
| | Ni | $d^8$ | BS(M) | –34.0 | –16.3 | –15.2 | 108% | –1.2 |
| $O^{2-}$ | Co | $d^7$ | BS(M) | 69.1 | –38.2 | –34.7 | 110% | –3.5 |
| | Ni | $d^8$ | BS(M) | 109.5 | –122.8 | –124.0 | 99% | 1.2 |
| | Eu | $f^6$ | BS(Cu) | –9.34 | 0.124 | 0.167 | 74% | –0.043 |
| | Gd | $f^7$ | BS(Cu) | –3.64 | –0.090 | –0.073 | 124% | –0.017 |
| | Tb | $f^8$ | BS(Cu) | –1.25 | –0.140 | –0.099 | 141% | –0.041 |
| | Dy | $f^9$ | BS(M) | –0.76 | –0.262 | –0.205 | 128% | –0.058 |

**b)** Silver clusters $[M_2AgF_8]^{2-}$ (M = Ni, Co), $[M_2AgF_7O]^{3-}$ (M = Ni, Co) and $[M_2AgF_7O]^{-}$ (M = Eu, Gd, Tb)

| Ligand | Metal | v.e. | GS | $J_{AgM}$ / cm$^{-1}$ | $J_{MM}$(Ag) / cm$^{-1}$ | $J_{MM}$(Cd) / cm$^{-1}$ | Ratio | Δ / cm$^{-1}$ |
|---|---|---|---|---|---|---|---|---|
| $F^-$ | Co | $d^7$ | BS(M) | 41.0 | –17.3 | –18.2 | 95% | 0.9 |
| | Ni | $d^8$ | BS(Ag) | –44.4 | –12.5 | –11.1 | 112% | –1.4 |
| $O^{2-}$ | Co | $d^7$ | HS* | 326.5 | –71.1 | –31.1 | 229% | –40.0 |
| | Ni | $d^8$ | HS | 409.1 | 1670.6 | –100.5 | –1662% | 1771.1 |
| | Eu | $f^6$ | BS(Ag) | –22.60 | –0.016 | 0.084 | –19% | –0.100 |
| | Gd | $f^7$ | BS(Ag) | –12.54 | –0.146 | –0.119 | 123% | –0.027 |
| | Tb | $f^8$ | BS(M) | –4.30 | –0.160 | –0.130 | 123% | –0.030 |
| | Dy | $f^9$ | BS(M) | –1.47 | –0.293 | –0.197 | 148% | –0.096 |

**c)** Copper clusters total energy and Löwdin spin populations of each orbital (specified with $m_l$ number) in 3*d* or 4*f* subshell

| $[Co_2CuF_8]^{2-}$ spin state | Energy / eV | Co1 d(z$^2$) | d(xz) | d(yz) | d(x$^2$–y$^2$) | d(xy) | Co2 d(z$^2$) | d(xz) | d(yz) | d(x$^2$–y$^2$) | d(xy) |
|---|---|---|---|---|---|---|---|---|---|---|---|
| Zn, BS | -147053.225602 | -0.91 | -0.08 | -0.40 | -0.77 | -0.56 | 0.92 | 0.08 | 0.36 | 0.76 | 0.59 |
| Zn, HS | -147053.217071 | 0.91 | 0.08 | 0.40 | 0.77 | 0.56 | 0.92 | 0.08 | 0.37 | 0.76 | 0.59 |
| Cu, HS | -143179.215293 | 0.90 | 0.09 | 0.40 | 0.78 | 0.56 | 0.92 | 0.08 | 0.37 | 0.76 | 0.59 |
| Cu, BS(Cu) | -143179.219900 | 0.91 | 0.08 | 0.40 | 0.77 | 0.56 | 0.92 | 0.08 | 0.37 | 0.76 | 0.59 |
| Cu, BS(M1) | -143179.226474 | -0.91 | -0.08 | -0.40 | -0.77 | -0.56 | 0.92 | 0.08 | 0.37 | 0.76 | 0.59 |
| Cu, BS(M2) | -143179.225781 | 0.90 | 0.08 | 0.40 | 0.78 | 0.56 | -0.92 | -0.08 | -0.37 | -0.76 | -0.59 |

| $[Ni_2CuF_8]^{2-}$ spin state | Energy / eV | Ni1 d(z$^2$) | d(xz) | d(yz) | d(x$^2$–y$^2$) | d(xy) | Ni2 d(z$^2$) | d(xz) | d(yz) | d(x$^2$–y$^2$) | d(xy) |
|---|---|---|---|---|---|---|---|---|---|---|---|
| Zn, BS | -154035.397390 | -0.47 | -0.02 | -0.39 | -0.52 | -0.36 | 0.46 | 0.05 | 0.36 | 0.73 | 0.17 |
| Zn, HS | -154035.393643 | 0.46 | 0.02 | 0.39 | 0.52 | 0.36 | 0.45 | 0.05 | 0.36 | 0.73 | 0.17 |
| Cu, HS | -150161.393814 | 0.46 | 0.02 | 0.39 | 0.52 | 0.37 | 0.45 | 0.05 | 0.37 | 0.73 | 0.17 |
| Cu, BS(Cu) | -150161.402199 | 0.46 | 0.02 | 0.39 | 0.52 | 0.36 | 0.45 | 0.05 | 0.36 | 0.73 | 0.17 |
| Cu, BS(M1) | -150161.401659 | -0.46 | -0.02 | -0.39 | -0.53 | -0.36 | 0.45 | 0.05 | 0.36 | 0.73 | 0.17 |
| Cu, BS(M2) | -150161.402425 | 0.46 | 0.02 | 0.39 | 0.52 | 0.37 | -0.45 | -0.05 | -0.36 | -0.73 | -0.17 |

| $[Co_2CuF_7O]^{3-}$ | Energy / eV | **Co1** | | | | | **Co2** | | | | |
|---|---|---|---|---|---|---|---|---|---|---|---|
| spin state | | $d(z^2)$ | d(xz) | d(yz) | $d(x^2–y^2)$ | d(xy) | $d(z^2)$ | d(xz) | d(yz) | $d(x^2–y^2)$ | d(xy) |
| Zn, BS | -146375.939795 | -0.88 | -0.16 | -0.50 | -0.70 | -0.47 | 0.89 | 0.18 | 0.49 | 0.67 | 0.49 |
| Zn, HS | -146375.920499 | 0.88 | 0.16 | 0.51 | 0.70 | 0.47 | 0.89 | 0.17 | 0.49 | 0.68 | 0.49 |
| Cu, HS | -142502.048229 | 0.89 | 0.16 | 0.51 | 0.70 | 0.47 | 0.89 | 0.17 | 0.49 | 0.68 | 0.49 |
| Cu, BS(Cu) | -142502.022638 | 0.88 | 0.16 | 0.51 | 0.70 | 0.47 | 0.89 | 0.17 | 0.49 | 0.68 | 0.50 |
| Cu, BS(M1) | -142502.057222 | -0.88 | -0.16 | -0.50 | -0.70 | -0.47 | 0.89 | 0.18 | 0.49 | 0.67 | 0.49 |
| Cu, BS(M2) | -142502.056138 | 0.88 | 0.17 | 0.50 | 0.70 | 0.47 | -0.89 | -0.17 | -0.49 | -0.67 | -0.49 |

| $[Ni_2CuF_7O]^{3-}$ | Energy / eV | **Ni1** | | | | | **Ni2** | | | | |
|---|---|---|---|---|---|---|---|---|---|---|---|
| spin state | | $d(z^2)$ | d(xz) | d(yz) | $d(x^2–y^2)$ | d(xy) | $d(z^2)$ | d(xz) | d(yz) | $d(x^2–y^2)$ | d(xy) |
| Zn, BS | -153358.081123 | -0.33 | -0.11 | -0.73 | -0.39 | -0.20 | 0.31 | 0.13 | 0.71 | 0.39 | 0.21 |
| Zn, HS | -153358.050511 | 0.33 | 0.11 | 0.73 | 0.39 | 0.20 | 0.31 | 0.13 | 0.72 | 0.39 | 0.21 |
| Cu, HS | -149484.182358 | 0.33 | 0.11 | 0.73 | 0.39 | 0.20 | 0.31 | 0.14 | 0.72 | 0.39 | 0.21 |
| Cu, BS(Cu) | -149484.155323 | 0.33 | 0.11 | 0.73 | 0.39 | 0.20 | 0.31 | 0.13 | 0.72 | 0.40 | 0.21 |
| Cu, BS(M1) | -149484.199021 | -0.33 | -0.11 | -0.73 | -0.39 | -0.20 | 0.31 | 0.14 | 0.71 | 0.39 | 0.21 |
| Cu, BS(M2) | -149484.199311 | 0.33 | 0.11 | 0.72 | 0.39 | 0.20 | -0.31 | -0.13 | -0.71 | -0.39 | -0.21 |

| $[Eu_2CuF_7O]^{-}$ | Energy / eV | **Eu1** | | | | | | | **Eu2** | | | | | | |
|---|---|---|---|---|---|---|---|---|---|---|---|---|---|---|---|
| spin state | | f0 | f+1 | f-1 | f+2 | f-2 | f+3 | f-3 | f0 | f+1 | f-1 | f+2 | f-2 | f+3 | f-3 |
| Zn, BS | -673148.050212 | -0.97 | -0.87 | -0.47 | -0.97 | -0.96 | -0.76 | -0.97 | 0.96 | 0.64 | 0.70 | 0.96 | 0.99 | 0.87 | 0.85 |
| Zn, HS | -673148.050583 | 0.97 | 0.87 | 0.47 | 0.97 | 0.97 | 0.75 | 0.97 | 0.96 | 0.64 | 0.70 | 0.96 | 0.99 | 0.88 | 0.84 |
| Cu, HS | -669273.986836 | 0.97 | 0.87 | 0.47 | 0.97 | 0.96 | 0.76 | 0.97 | 0.96 | 0.64 | 0.70 | 0.96 | 0.99 | 0.88 | 0.84 |
| Cu, BS(Cu) | -669273.993752 | 0.97 | 0.87 | 0.47 | 0.97 | 0.96 | 0.76 | 0.96 | 0.96 | 0.64 | 0.70 | 0.96 | 0.99 | 0.87 | 0.85 |
| Cu, BS(M1) | -669273.990023 | -0.97 | -0.87 | -0.47 | -0.97 | -0.96 | -0.76 | -0.96 | 0.96 | 0.64 | 0.70 | 0.96 | 0.99 | 0.87 | 0.85 |
| Cu, BS(M2) | -669273.990016 | 0.97 | 0.87 | 0.47 | 0.97 | 0.96 | 0.76 | 0.96 | -0.96 | -0.64 | -0.70 | -0.96 | -0.99 | -0.87 | -0.85 |

| $[Gd_2CuF_7O]^{-}$ | Energy / eV | **Gd1** | | | | | | | **Gd2** | | | | | | |
|---|---|---|---|---|---|---|---|---|---|---|---|---|---|---|---|
| spin state | | f0 | f+1 | f-1 | f+2 | f-2 | f+3 | f-3 | f0 | f+1 | f-1 | f+2 | f-2 | f+3 | f-3 |
| Zn, BS | -697339.551093 | -0.99 | -0.99 | -0.99 | -0.99 | -0.99 | -0.99 | -0.99 | 0.99 | 0.99 | 0.99 | 0.99 | 0.99 | 0.99 | 0.99 |
| Zn, HS | -697339.550872 | 0.99 | 0.99 | 0.99 | 0.99 | 0.99 | 0.99 | 0.99 | 0.99 | 0.99 | 0.99 | 0.99 | 0.99 | 0.99 | 0.99 |
| Cu, HS | -693465.478415 | 0.99 | 0.99 | 0.99 | 0.99 | 0.99 | 0.99 | 0.99 | 0.99 | 0.99 | 0.99 | 0.99 | 0.99 | 0.99 | 0.99 |
| Cu, BS(Cu) | -693465.481561 | 0.99 | 0.99 | 0.99 | 0.99 | 0.99 | 0.99 | 0.99 | 0.99 | 0.99 | 0.99 | 0.99 | 0.99 | 0.99 | 0.99 |
| Cu, BS(M1) | -693465.480262 | -0.99 | -0.99 | -0.99 | -0.99 | -0.99 | -0.99 | -0.99 | 0.99 | 0.99 | 0.99 | 0.99 | 0.99 | 0.99 | 0.99 |
| Cu, BS(M2) | -693465.480260 | 0.99 | 0.99 | 0.99 | 0.99 | 0.99 | 0.99 | 0.99 | -0.99 | -0.99 | -0.99 | -0.99 | -0.99 | -0.99 | -0.99 |

| $[Tb_2CuF_7O]^{-}$ | Energy / eV | **Tb1** | | | | | | | **Tb2** | | | | | | |
|---|---|---|---|---|---|---|---|---|---|---|---|---|---|---|---|
| spin state | | f0 | f+1 | f-1 | f+2 | f-2 | f+3 | f-3 | f0 | f+1 | f-1 | f+2 | f-2 | f+3 | f-3 |
| Zn, BS | -722134.249344 | -0.97 | -0.86 | -0.56 | -0.94 | -0.96 | -0.65 | -0.98 | 0.99 | 0.97 | 0.40 | 0.99 | 0.98 | 0.96 | 0.65 |
| Zn, HS | -722134.249124 | 0.97 | 0.86 | 0.56 | 0.95 | 0.96 | 0.65 | 0.98 | 0.99 | 0.97 | 0.40 | 0.99 | 0.98 | 0.96 | 0.65 |
| Cu, HS | -718260.171907 | 0.97 | 0.88 | 0.57 | 0.93 | 0.94 | 0.66 | 0.98 | 0.99 | 0.97 | 0.42 | 0.99 | 0.95 | 0.96 | 0.65 |
| Cu, BS(Cu) | -718260.172835 | 0.96 | 0.88 | 0.58 | 0.92 | 0.94 | 0.66 | 0.98 | 0.99 | 0.97 | 0.42 | 0.99 | 0.95 | 0.95 | 0.66 |
| Cu, BS(M1) | -718260.172676 | 0.97 | 0.88 | 0.57 | 0.93 | 0.94 | 0.66 | 0.98 | -0.99 | -0.97 | -0.42 | -0.99 | -0.95 | -0.95 | -0.66 |
| Cu, BS(M2) | -718260.172688 | -0.97 | -0.88 | -0.58 | -0.93 | -0.94 | -0.66 | -0.98 | 0.99 | 0.97 | 0.42 | 0.99 | 0.95 | 0.96 | 0.65 |

| $[Dy_2CuF_7O]^{-}$ | Energy / eV | **Dy1** | | | | | | | **Dy2** | | | | | | |
|---|---|---|---|---|---|---|---|---|---|---|---|---|---|---|---|
| spin state | | f0 | f+1 | f-1 | f+2 | f-2 | f+3 | f-3 | f0 | f+1 | f-1 | f+2 | f-2 | f+3 | f-3 |
| Zn, BS | -747581.408492 | -0.94 | -0.28 | -0.54 | -0.96 | -0.97 | -0.28 | -0.95 | 0.94 | 0.52 | 0.32 | 0.98 | 0.94 | 0.82 | 0.42 |
| Zn, HS | -747581.408176 | 0.95 | 0.28 | 0.54 | 0.96 | 0.97 | 0.28 | 0.95 | 0.94 | 0.52 | 0.32 | 0.98 | 0.94 | 0.82 | 0.42 |
| Cu, HS | -743707.333236 | 0.93 | 0.29 | 0.54 | 0.96 | 0.97 | 0.28 | 0.95 | 0.94 | 0.52 | 0.32 | 0.98 | 0.94 | 0.82 | 0.42 |
| Cu, BS(Cu) | -743707.333705 | 0.94 | 0.28 | 0.54 | 0.96 | 0.97 | 0.28 | 0.95 | 0.93 | 0.53 | 0.32 | 0.98 | 0.93 | 0.81 | 0.42 |
| Cu, BS(M1) | -743707.333877 | -0.93 | -0.29 | -0.54 | -0.96 | -0.97 | -0.28 | -0.95 | 0.94 | 0.52 | 0.32 | 0.98 | 0.94 | 0.82 | 0.42 |
| Cu, BS(M2) | -743707.333874 | 0.93 | 0.29 | 0.54 | 0.96 | 0.97 | 0.28 | 0.95 | -0.94 | -0.52 | -0.31 | -0.98 | -0.94 | -0.82 | -0.42 |

**Table S6.** Computed data for TEMPO analogues of molecules presented in **Figure 1** in the main article, of a formula $[M_2F_4X{\cdot}TEMPO]$. All systems were ionically relaxed in their GS configuration. XYZ are provided in the 6th section of SI. The reference system used is a substitution of neutral TEMPO with an $O^{2-}$ ligand. Ratio is defined as Ratio=$J_{MM}$(TEMPO)/$J_{MM}$(Ref.), while difference as Δ=$J_{MM}$(TEMPO)–$J_{MM}$(Ref.). Note that a considerable difference between X = $F^-$ and $O^{2-}$ for a $Gd^{3+}$ cation arises from the geometry change between the two, i.e. with X = $F^-$ we notice a rotation of TEMPO ring with respect to $Gd_2XO$ plane, which causes rotation of 2p hole on O(TEMPO) out of Gd-O(TEMPO)-Gd plane. The $J_{MM}$ with a TEMPO radical for a Ni system could not be obtained due to orbital configuration changes between a different spin states (i.e. for BS(M1) and BS(M2) states).

| Ligand | Metal | v.e. | GS | $J_{TEMPO\text{-}M}$ / $cm^{-1}$ | $J_{MM}$(TEMPO) / $cm^{-1}$ | $J_{MM}$(Ref.) / $cm^{-1}$ | Ratio | Δ / $cm^{-1}$ |
|---|---|---|---|---|---|---|---|---|
| $F^-$ | Ni | $d^8$ | HS | 228.8 | - | 224.9 | - | - |
| | Gd | $f^7$ | BS(M) | 1.0 | –0.14 | –0.10 | 145% | –0.04 |
| $O^{2-}$ | Gd | $f^7$ | HS | 11.5 | –0.39 | 0.54 | –73% | –0.93 |

## 3. Oxide [$AgM_2F_7O$] clusters

**Table S7.** Total energy and Löwdin spin populations of each orbital (specified with $m_l$ number) in 3*d* or 4*f* subshell for each 3d TM or Ln cation in the clusters of [$AgM_2F_7O$] type.

| State | Energy / eV | Co1 | | | | | Co2 | | | | |
|---|---|---|---|---|---|---|---|---|---|---|---|
| | | d(z$^2$) | d(xz) | d(yz) | d(x$^2$–y$^2$) | d(xy) | d(z$^2$) | d(xz) | d(yz) | d(x$^2$–y$^2$) | d(xy) |
| Cd, BS | -251255.386310 | -0.85 | -0.19 | -0.51 | -0.76 | -0.42 | 0.85 | 0.21 | 0.49 | 0.73 | 0.44 |
| Cd, HS | -251255.369040 | 0.86 | 0.18 | 0.51 | 0.76 | 0.42 | 0.85 | 0.20 | 0.49 | 0.74 | 0.44 |
| Ag, HS | -243639.667685 | 0.85 | 0.24 | 0.50 | 0.76 | 0.43 | 0.84 | 0.26 | 0.48 | 0.74 | 0.45 |
| Ag, BS(Ag) | -243639.546773 | 0.86 | 0.16 | 0.50 | 0.76 | 0.43 | 0.85 | 0.19 | 0.48 | 0.74 | 0.46 |
| Ag, BS(M1) | -243639.646739 | -0.85 | -0.17 | -0.49 | -0.75 | -0.44 | 0.83 | 0.27 | 0.49 | 0.75 | 0.44 |
| Ag, BS(M2) | -243639.646743 | 0.84 | 0.24 | 0.51 | 0.77 | 0.41 | -0.84 | -0.20 | -0.47 | -0.73 | -0.46 |

| State | Energy / eV | Ni1 | | | | | Ni2 | | | | |
|---|---|---|---|---|---|---|---|---|---|---|---|
| | | d(z$^2$) | d(xz) | d(yz) | d(x$^2$–y$^2$) | d(xy) | d(z$^2$) | d(xz) | d(yz) | d(x$^2$–y$^2$) | d(xy) |
| Cd, BS | -258237.494750 | -0.27 | -0.18 | -0.69 | -0.39 | -0.24 | 0.24 | 0.22 | 0.67 | 0.39 | 0.24 |
| Cd, HS | -258237.469930 | 0.26 | 0.18 | 0.70 | 0.39 | 0.24 | 0.24 | 0.22 | 0.68 | 0.39 | 0.24 |
| Ag, HS | -250621.714866 | 0.26 | 0.20 | 0.68 | 0.40 | 0.25 | 0.24 | 0.24 | 0.66 | 0.40 | 0.26 |
| Ag, BS(Ag) | -250621.613866 | 0.24 | 0.18 | 0.69 | 0.40 | 0.25 | 0.22 | 0.22 | 0.67 | 0.40 | 0.25 |
| Ag, BS(M1) | -250621.251860 | -0.25 | -0.17 | -0.68 | -0.40 | -0.25 | 0.24 | 0.24 | 0.65 | 0.40 | 0.25 |
| Ag, BS(M2) | -250621.251886 | 0.26 | 0.20 | 0.67 | 0.40 | 0.25 | -0.22 | -0.21 | -0.66 | -0.40 | -0.25 |

| State | Energy / eV | Cu1 | | | | | Cu2 | | | | |
|---|---|---|---|---|---|---|---|---|---|---|---|
| | | d(z$^2$) | d(xz) | d(yz) | d(x$^2$–y$^2$) | d(xy) | d(z$^2$) | d(xz) | d(yz) | d(x$^2$–y$^2$) | d(xy) |
| Cd, BS | -265598.697108 | -0.50 | -0.06 | -0.02 | -0.03 | -0.13 | 0.48 | 0.08 | 0.01 | 0.04 | 0.13 |
| Cd, HS | -265598.673592 | 0.51 | 0.05 | 0.02 | 0.04 | 0.13 | 0.49 | 0.07 | 0.02 | 0.04 | 0.13 |
| Ag, HS | -257982.945014 | 0.49 | 0.07 | 0.02 | 0.04 | 0.14 | 0.47 | 0.09 | 0.02 | 0.04 | 0.14 |
| Ag, BS(Ag) | -257982.976526 | 0.50 | 0.05 | 0.02 | 0.04 | 0.14 | 0.48 | 0.07 | 0.02 | 0.04 | 0.13 |
| Ag, BS(M1) | -257982.995891 | -0.49 | -0.06 | -0.02 | -0.03 | -0.13 | 0.47 | 0.09 | 0.02 | 0.04 | 0.13 |
| Ag, BS(M2) | -257983.000794 | 0.49 | 0.07 | 0.02 | 0.03 | 0.13 | -0.47 | -0.08 | -0.02 | -0.04 | -0.13 |

| State | Energy / eV | Sm1 | | | | | | | Sm2 | | | | | | |
|---|---|---|---|---|---|---|---|---|---|---|---|---|---|---|---|
| | | f0 | f+1 | f-1 | f+2 | f-2 | f+3 | f-3 | f0 | f+1 | f-1 | f+2 | f-2 | f+3 | f-3 |
| Cd, BS | -754456.534500 | -0.53 | -0.65 | -0.84 | -0.67 | -0.83 | -0.79 | -0.67 | 0.66 | 0.45 | 0.88 | 0.65 | 0.92 | 0.54 | 0.89 |
| Cd, HS | -754456.532929 | 0.53 | 0.65 | 0.84 | 0.67 | 0.82 | 0.78 | 0.68 | 0.66 | 0.45 | 0.87 | 0.65 | 0.92 | 0.53 | 0.89 |
| Ag, HS | -746840.437068 | 0.53 | 0.65 | 0.84 | 0.67 | 0.82 | 0.78 | 0.68 | 0.66 | 0.45 | 0.87 | 0.65 | 0.92 | 0.53 | 0.89 |
| Ag, BS(Ag) | -746840.455552 | 0.52 | 0.66 | 0.83 | 0.66 | 0.84 | 0.79 | 0.67 | 0.64 | 0.47 | 0.86 | 0.65 | 0.93 | 0.51 | 0.91 |
| Ag, BS(M1) | -746840.448028 | -0.52 | -0.65 | -0.84 | -0.66 | -0.84 | -0.79 | -0.67 | 0.66 | 0.45 | 0.88 | 0.65 | 0.92 | 0.54 | 0.89 |
| Ag, BS(M2) | -746840.447903 | 0.54 | 0.65 | 0.84 | 0.67 | 0.82 | 0.79 | 0.68 | -0.63 | -0.49 | -0.87 | -0.64 | -0.93 | -0.51 | -0.91 |

| State | Energy / eV | Eu1 | | | | | | | Eu2 | | | | | | |
|---|---|---|---|---|---|---|---|---|---|---|---|---|---|---|---|
| | | f0 | f+1 | f-1 | f+2 | f-2 | f+3 | f-3 | f0 | f+1 | f-1 | f+2 | f-2 | f+3 | f-3 |
| Cd, BS | -778027.349127 | -0.78 | -0.89 | -0.66 | -0.98 | -0.98 | -0.72 | -0.96 | 0.96 | 0.63 | 0.83 | 0.83 | 0.98 | 0.89 | 0.87 |
| Cd, HS | -778027.349314 | 0.78 | 0.89 | 0.66 | 0.98 | 0.98 | 0.71 | 0.96 | 0.96 | 0.63 | 0.83 | 0.83 | 0.98 | 0.89 | 0.86 |
| Ag, HS | -770411.260243 | 0.78 | 0.88 | 0.66 | 0.98 | 0.98 | 0.72 | 0.97 | 0.96 | 0.63 | 0.83 | 0.83 | 0.98 | 0.89 | 0.87 |
| Ag, BS(Ag) | -770411.276986 | 0.78 | 0.89 | 0.66 | 0.98 | 0.98 | 0.72 | 0.95 | 0.96 | 0.63 | 0.83 | 0.83 | 0.98 | 0.88 | 0.87 |
| Ag, BS(M1) | -770411.268655 | -0.78 | -0.89 | -0.66 | -0.98 | -0.98 | -0.72 | -0.95 | 0.96 | 0.63 | 0.83 | 0.83 | 0.98 | 0.89 | 0.87 |
| Ag, BS(M2) | -770411.268644 | 0.78 | 0.89 | 0.66 | 0.98 | 0.98 | 0.72 | 0.96 | -0.95 | -0.63 | -0.83 | -0.83 | -0.98 | -0.87 | -0.88 |

| State | Energy / eV | Gd1 | | | | | | | Gd2 | | | | | | |
|---|---|---|---|---|---|---|---|---|---|---|---|---|---|---|---|
| | | f0 | f+1 | f-1 | f+2 | f-2 | f+3 | f-3 | f0 | f+1 | f-1 | f+2 | f-2 | f+3 | f-3 |
| Cd, BS | -802215.882009 | -0.99 | -0.99 | -0.99 | -0.99 | -0.99 | -0.99 | -0.99 | 0.99 | 0.99 | 0.99 | 0.99 | 0.99 | 0.99 | 0.99 |
| Cd, HS | -802215.881650 | 0.99 | 0.99 | 0.99 | 0.99 | 0.99 | 0.99 | 0.99 | 0.99 | 0.99 | 0.99 | 0.99 | 0.99 | 0.99 | 0.99 |
| Ag, HS | -794597.616837 | 0.99 | 0.99 | 0.99 | 0.99 | 0.99 | 0.99 | 0.99 | 0.99 | 0.99 | 0.99 | 0.99 | 0.99 | 0.99 | 0.99 |
| Ag, BS(Ag) | -794597.627676 | 0.99 | 0.99 | 0.99 | 0.99 | 0.99 | 0.99 | 0.99 | 0.99 | 0.99 | 0.99 | 0.99 | 0.99 | 0.99 | 0.99 |
| Ag, BS(M1) | -794597.622703 | -0.99 | -0.99 | -0.99 | -0.99 | -0.99 | -0.99 | -0.99 | 0.99 | 0.99 | 0.99 | 0.99 | 0.99 | 0.99 | 0.99 |
| Ag, BS(M2) | -794597.622693 | 0.99 | 0.99 | 0.99 | 0.99 | 0.99 | 0.99 | 0.99 | -0.99 | -0.99 | -0.99 | -0.99 | -0.99 | -0.99 | -0.99 |

| State | Energy / eV | Tb1 | | | | | | | Tb2 | | | | | | |
|---|---|---|---|---|---|---|---|---|---|---|---|---|---|---|---|
| | | f0 | f+1 | f-1 | f+2 | f-2 | f+3 | f-3 | f0 | f+1 | f-1 | f+2 | f-2 | f+3 | f-3 |
| Cd, BS | -827019.084945 | -0.99 | -0.81 | -0.55 | -0.99 | -0.99 | -0.61 | -0.99 | 0.94 | 0.52 | 0.91 | 0.98 | 0.99 | 0.83 | 0.75 |
| Cd, HS | -827019.084656 | 0.99 | 0.81 | 0.55 | 0.99 | 0.99 | 0.61 | 0.99 | 0.94 | 0.52 | 0.91 | 0.99 | 0.99 | 0.83 | 0.75 |
| Ag, HS | -819402.974393 | 0.98 | 0.84 | 0.55 | 0.98 | 0.98 | 0.63 | 0.98 | 0.94 | 0.51 | 0.92 | 0.98 | 0.99 | 0.79 | 0.80 |
| Ag, BS(Ag) | -819402.977579 | 0.98 | 0.82 | 0.57 | 0.97 | 0.98 | 0.62 | 0.99 | 0.94 | 0.52 | 0.91 | 0.99 | 0.99 | 0.82 | 0.75 |
| Ag, BS(M1) | -819402.978848 | 0.99 | 0.84 | 0.54 | 0.98 | 0.98 | 0.62 | 0.98 | -0.94 | -0.52 | -0.91 | -0.99 | -0.99 | -0.82 | -0.75 |
| Ag, BS(M2) | -819402.973835 | -0.98 | -0.82 | -0.56 | -0.97 | -0.98 | -0.62 | -0.99 | 0.94 | 0.51 | 0.92 | 0.98 | 0.99 | 0.79 | 0.80 |

| State | Energy / eV | Dy1 | | | | | | | Dy2 | | | | | | |
|---|---|---|---|---|---|---|---|---|---|---|---|---|---|---|---|
| | | f0 | f+1 | f-1 | f+2 | f-2 | f+3 | f-3 | f0 | f+1 | f-1 | f+2 | f-2 | f+3 | f-3 |
| Cd, BS | -852460.708168 | -0.94 | -0.25 | -0.56 | -0.96 | -0.98 | -0.30 | -0.94 | 0.93 | 0.49 | 0.34 | 0.98 | 0.94 | 0.75 | 0.49 |
| Cd, HS | -852460.707864 | 0.95 | 0.25 | 0.56 | 0.96 | 0.98 | 0.30 | 0.94 | 0.93 | 0.49 | 0.34 | 0.98 | 0.94 | 0.75 | 0.50 |
| Ag, HS | -844844.595358 | 0.94 | 0.26 | 0.56 | 0.96 | 0.97 | 0.30 | 0.94 | 0.93 | 0.49 | 0.34 | 0.98 | 0.94 | 0.77 | 0.48 |
| Ag, BS(Ag) | -844844.596262 | 0.94 | 0.25 | 0.56 | 0.96 | 0.98 | 0.30 | 0.94 | 0.93 | 0.49 | 0.34 | 0.98 | 0.94 | 0.75 | 0.49 |
| Ag, BS(M1) | -844844.596332 | -0.94 | -0.25 | -0.56 | -0.96 | -0.98 | -0.30 | -0.94 | 0.93 | 0.49 | 0.34 | 0.98 | 0.94 | 0.77 | 0.48 |
| Ag, BS(M2) | -844844.596193 | 0.94 | 0.26 | 0.56 | 0.96 | 0.97 | 0.30 | 0.94 | -0.93 | -0.49 | -0.34 | -0.98 | -0.94 | -0.75 | -0.49 |

| State | Energy / eV | Ho1 | | | | | | | Ho2 | | | | | | |
|---|---|---|---|---|---|---|---|---|---|---|---|---|---|---|---|
| | | f0 | f+1 | f-1 | f+2 | f-2 | f+3 | f-3 | f0 | f+1 | f-1 | f+2 | f-2 | f+3 | f-3 |
| Cd, BS | -878550.475829 | 0.69 | 0.24 | 0.72 | 0.66 | 0.53 | 0.19 | 0.93 | -0.67 | -0.13 | -0.84 | -0.25 | -0.93 | -0.87 | -0.25 |
| Cd, HS | -878550.475715 | 0.69 | 0.24 | 0.72 | 0.67 | 0.52 | 0.19 | 0.93 | 0.67 | 0.13 | 0.84 | 0.25 | 0.93 | 0.87 | 0.25 |
| Ag, HS | -870934.360557 | 0.69 | 0.21 | 0.75 | 0.58 | 0.60 | 0.25 | 0.87 | 0.68 | 0.14 | 0.84 | 0.22 | 0.95 | 0.92 | 0.20 |
| Ag, BS(Ag) | -870934.359151 | 0.69 | 0.23 | 0.73 | 0.65 | 0.54 | 0.19 | 0.92 | 0.68 | 0.13 | 0.84 | 0.24 | 0.94 | 0.88 | 0.23 |
| Ag, BS(M1) | -870934.360126 | -0.69 | -0.25 | -0.71 | -0.63 | -0.56 | -0.19 | -0.92 | 0.68 | 0.14 | 0.84 | 0.22 | 0.95 | 0.93 | 0.20 |
| Ag, BS(M2) | -870934.360136 | 0.69 | 0.21 | 0.75 | 0.57 | 0.61 | 0.25 | 0.87 | -0.68 | -0.13 | -0.84 | -0.24 | -0.94 | -0.89 | -0.23 |

| State | Energy / eV | Er1 | | | | | | | Er2 | | | | | | |
|---|---|---|---|---|---|---|---|---|---|---|---|---|---|---|---|
| | | f0 | f+1 | f-1 | f+2 | f-2 | f+3 | f-3 | f0 | f+1 | f-1 | f+2 | f-2 | f+3 | f-3 |
| Cd, BS | -905298.013061 | -0.24 | -0.34 | -0.78 | -0.23 | -0.52 | -0.03 | -0.83 | 0.27 | 0.55 | 0.52 | 0.07 | 0.69 | 0.58 | 0.27 |
| Cd, HS | -905298.012988 | 0.24 | 0.34 | 0.78 | 0.24 | 0.51 | 0.03 | 0.83 | 0.27 | 0.55 | 0.52 | 0.07 | 0.69 | 0.58 | 0.28 |
| Ag, HS | -897673.739213 | 0.23 | 0.35 | 0.77 | 0.21 | 0.53 | 0.03 | 0.83 | 0.27 | 0.56 | 0.51 | 0.08 | 0.69 | 0.61 | 0.24 |
| Ag, BS(Ag) | -897673.742544 | 0.24 | 0.34 | 0.78 | 0.24 | 0.51 | 0.03 | 0.82 | 0.27 | 0.55 | 0.52 | 0.07 | 0.69 | 0.57 | 0.28 |
| Ag, BS(M1) | -897673.741049 | -0.24 | -0.34 | -0.78 | -0.24 | -0.51 | -0.03 | -0.82 | 0.27 | 0.56 | 0.51 | 0.08 | 0.69 | 0.62 | 0.24 |
| Ag, BS(M2) | -897673.741028 | 0.23 | 0.35 | 0.77 | 0.21 | 0.54 | 0.03 | 0.83 | -0.27 | -0.55 | -0.52 | -0.07 | -0.69 | -0.57 | -0.28 |

| State | Energy / eV | Tm1 | | | | | | | Tm2 | | | | | | |
|---|---|---|---|---|---|---|---|---|---|---|---|---|---|---|---|
| | | f0 | f+1 | f-1 | f+2 | f-2 | f+3 | f-3 | f0 | f+1 | f-1 | f+2 | f-2 | f+3 | f-3 |
| Cd, BS | -932714.187618 | -0.06 | -0.41 | -0.80 | -0.05 | -0.04 | -0.21 | -0.41 | 0.04 | 0.46 | 0.78 | 0.05 | 0.01 | 0.47 | 0.16 |
| Cd, HS | -932714.187414 | 0.03 | 0.44 | 0.81 | 0.03 | 0.02 | 0.22 | 0.41 | 0.04 | 0.45 | 0.78 | 0.05 | 0.01 | 0.47 | 0.15 |
| Ag, HS | -925098.065351 | 0.03 | 0.43 | 0.82 | 0.03 | 0.02 | 0.22 | 0.41 | 0.04 | 0.44 | 0.80 | 0.05 | 0.01 | 0.48 | 0.14 |
| Ag, BS(Ag) | -925098.059992 | 0.03 | 0.44 | 0.82 | 0.03 | 0.02 | 0.22 | 0.40 | 0.04 | 0.45 | 0.78 | 0.05 | 0.01 | 0.47 | 0.15 |
| Ag, BS(M1) | -925098.062883 | -0.02 | -0.46 | -0.82 | -0.02 | -0.02 | -0.22 | -0.41 | 0.04 | 0.44 | 0.80 | 0.05 | 0.01 | 0.48 | 0.14 |
| Ag, BS(M2) | -925098.062901 | 0.04 | 0.42 | 0.82 | 0.03 | 0.03 | 0.22 | 0.41 | -0.03 | -0.48 | -0.78 | -0.04 | -0.01 | -0.47 | -0.15 |

| State | Energy / eV | Yb1 | | | | | | | Yb2 | | | | | | |
|---|---|---|---|---|---|---|---|---|---|---|---|---|---|---|---|
| | | f0 | f+1 | f-1 | f+2 | f-2 | f+3 | f-3 | f0 | f+1 | f-1 | f+2 | f-2 | f+3 | f-3 |
| Cd, BS | -960812.676145 | -0.02 | -0.13 | -0.49 | -0.02 | -0.02 | -0.30 | 0.00 | 0.03 | 0.33 | 0.27 | 0.03 | 0.00 | 0.09 | 0.22 |
| Cd, HS | -960812.676179 | 0.02 | 0.13 | 0.49 | 0.02 | 0.02 | 0.30 | 0.00 | 0.03 | 0.33 | 0.27 | 0.03 | 0.00 | 0.09 | 0.22 |
| Ag, HS | -953196.552443 | 0.02 | 0.12 | 0.50 | 0.02 | 0.02 | 0.29 | 0.01 | 0.03 | 0.32 | 0.29 | 0.03 | 0.00 | 0.11 | 0.19 |
| Ag, BS(Ag) | -953196.550022 | 0.02 | 0.13 | 0.49 | 0.02 | 0.02 | 0.29 | 0.00 | 0.03 | 0.34 | 0.28 | 0.03 | 0.00 | 0.09 | 0.21 |
| Ag, BS(M1) | -953196.551208 | -0.02 | -0.12 | -0.49 | -0.02 | -0.02 | -0.29 | 0.00 | 0.03 | 0.32 | 0.29 | 0.03 | 0.00 | 0.11 | 0.19 |
| Ag, BS(M2) | -953196.551217 | 0.02 | 0.12 | 0.50 | 0.02 | 0.02 | 0.29 | 0.02 | -0.03 | -0.34 | -0.28 | -0.03 | 0.00 | -0.09 | -0.21 |

**Table S8.** Total energy and Löwdin spin populations of each orbital (specified with $m_l$ number) in 4*s*, 4*p* and 3*d* subshells for the silver cation in the clusters of $[AgM_2F_7O]^-$ type.

| State | Energy / eV | Co | | | | | | | | |
|---|---|---|---|---|---|---|---|---|---|---|
| | | s | $p_z$ | $p_x$ | $p_y$ | d($z^2$) | d(xz) | d(yz) | d($x^2$–$y^2$) | d(xy) |
| Ag, HS | -243639.667685 | 0.030 | 0.000 | -0.002 | -0.004 | 0.227 | 0.001 | 0.007 | 0.123 | 0.000 |
| Ag, BS(Ag) | -243639.546773 | -0.027 | 0.009 | 0.002 | 0.000 | -0.258 | 0.001 | -0.008 | -0.144 | 0.000 |
| Ag, BS(M1) | -243639.646739 | 0.029 | -0.004 | -0.002 | -0.002 | 0.234 | 0.000 | 0.008 | 0.130 | 0.000 |
| Ag, BS(M2) | -243639.646743 | 0.029 | -0.004 | -0.002 | -0.002 | 0.234 | 0.001 | 0.008 | 0.129 | 0.000 |

| State | Energy / eV | Ni | | | | | | | | |
|---|---|---|---|---|---|---|---|---|---|---|
| | | s | $p_z$ | $p_x$ | $p_y$ | d($z^2$) | d(xz) | d(yz) | d($x^2$–$y^2$) | d(xy) |
| Ag, HS | -250621.714866 | 0.031 | -0.001 | -0.003 | -0.004 | 0.239 | 0.000 | 0.000 | 0.095 | 0.034 |
| Ag, BS(Ag) | -250621.613866 | -0.030 | 0.008 | 0.002 | 0.001 | -0.263 | -0.001 | 0.000 | -0.103 | -0.037 |
| Ag, BS(M1) | -250621.251860 | 0.030 | -0.005 | -0.003 | -0.002 | 0.250 | 0.001 | 0.000 | 0.098 | 0.036 |
| Ag, BS(M2) | -250621.251886 | 0.030 | -0.005 | -0.003 | -0.002 | 0.250 | 0.000 | 0.000 | 0.099 | 0.036 |

| State | Energy / eV | Cu | | | | | | | | |
|---|---|---|---|---|---|---|---|---|---|---|
| | | s | $p_z$ | $p_x$ | $p_y$ | d($z^2$) | d(xz) | d(yz) | d($x^2$–$y^2$) | d(xy) |
| Ag, HS | -257982.945014 | 0.012 | -0.002 | -0.003 | -0.004 | 0.297 | 0.000 | 0.001 | 0.136 | 0.000 |
| Ag, BS(Ag) | -257982.976526 | -0.015 | 0.011 | 0.002 | 0.001 | -0.293 | 0.000 | -0.001 | -0.135 | 0.000 |
| Ag, BS(M1) | -257982.995891 | 0.013 | -0.006 | -0.003 | -0.003 | 0.293 | 0.000 | 0.001 | 0.135 | 0.000 |
| Ag, BS(M2) | -257983.000794 | 0.013 | -0.006 | -0.003 | -0.003 | 0.292 | 0.001 | 0.001 | 0.134 | 0.000 |

| State | Energy / eV | Sm | | | | | | | | |
|---|---|---|---|---|---|---|---|---|---|---|
| | | s | $p_z$ | $p_x$ | $p_y$ | d($z^2$) | d(xz) | d(yz) | d($x^2$–$y^2$) | d(xy) |
| Ag, HS | -746840.437068 | 0.029 | -0.001 | -0.005 | -0.005 | 0.003 | 0.020 | 0.069 | 0.337 | 0.075 |
| Ag, BS(Ag) | -746840.455552 | -0.029 | 0.002 | 0.004 | 0.005 | -0.003 | -0.019 | -0.068 | -0.335 | -0.074 |
| Ag, BS(M1) | -746840.448028 | 0.029 | -0.002 | -0.005 | -0.005 | 0.003 | 0.019 | 0.069 | 0.336 | 0.075 |
| Ag, BS(M2) | -746840.447903 | 0.029 | -0.002 | -0.005 | -0.005 | 0.003 | 0.019 | 0.069 | 0.336 | 0.075 |

| State | Energy / eV | Eu | | | | | | | | |
|---|---|---|---|---|---|---|---|---|---|---|
| | | s | $p_z$ | $p_x$ | $p_y$ | d($z^2$) | d(xz) | d(yz) | d($x^2$–$y^2$) | d(xy) |
| Ag, HS | -770411.260243 | 0.028 | -0.001 | -0.005 | -0.005 | 0.008 | 0.010 | 0.028 | 0.386 | 0.070 |
| Ag, BS(Ag) | -770411.276986 | -0.029 | 0.001 | 0.004 | 0.005 | -0.008 | -0.010 | -0.027 | -0.384 | -0.071 |
| Ag, BS(M1) | -770411.268655 | 0.028 | -0.001 | -0.005 | -0.005 | 0.008 | 0.010 | 0.028 | 0.385 | 0.070 |
| Ag, BS(M2) | -770411.268644 | 0.028 | -0.001 | -0.005 | -0.005 | 0.008 | 0.010 | 0.028 | 0.384 | 0.071 |

| State | Energy / eV | Gd | | | | | | | | |
|---|---|---|---|---|---|---|---|---|---|---|
| | | s | $p_z$ | $p_x$ | $p_y$ | d($z^2$) | d(xz) | d(yz) | d($x^2$–$y^2$) | d(xy) |
| Ag, HS | -794597.616837 | 0.031 | -0.001 | -0.005 | -0.005 | 0.015 | 0.004 | 0.001 | 0.418 | 0.066 |
| Ag, BS(Ag) | -794597.627676 | -0.031 | 0.001 | 0.004 | 0.006 | -0.015 | -0.004 | 0.000 | -0.416 | -0.064 |
| Ag, BS(M1) | -794597.622703 | 0.031 | -0.001 | -0.004 | -0.006 | 0.015 | 0.004 | 0.000 | 0.417 | 0.064 |
| Ag, BS(M2) | -794597.622693 | 0.031 | -0.001 | -0.005 | -0.005 | 0.015 | 0.004 | 0.000 | 0.417 | 0.065 |

| State | Energy / eV | Tb | | | | | | | | |
|---|---|---|---|---|---|---|---|---|---|---|
| | | s | $p_z$ | $p_x$ | $p_y$ | d($z^2$) | d(xz) | d(yz) | d($x^2$–$y^2$) | d(xy) |
| Ag, HS | -819402.974393 | 0.032 | -0.001 | -0.005 | -0.005 | 0.015 | 0.004 | 0.001 | 0.407 | 0.073 |
| Ag, BS(Ag) | -819402.977579 | -0.032 | 0.001 | 0.004 | 0.006 | -0.015 | -0.004 | 0.000 | -0.410 | -0.071 |
| Ag, BS(M1) | -819402.978848 | 0.032 | -0.001 | -0.005 | -0.005 | 0.015 | 0.004 | 0.000 | 0.407 | 0.072 |
| Ag, BS(M2) | -819402.973835 | 0.032 | -0.001 | -0.004 | -0.005 | 0.015 | 0.004 | 0.000 | 0.410 | 0.071 |

| State | Energy / eV | Dy s | Dy $p_z$ | Dy $p_x$ | Dy $p_y$ | Dy $d(z^2)$ | Dy d(xz) | Dy d(yz) | Dy $d(x^2-y^2)$ | Dy d(xy) |
|---|---|---|---|---|---|---|---|---|---|---|
| Ag, HS | -844844.595358 | 0.032 | -0.001 | -0.005 | -0.005 | 0.015 | 0.004 | 0.001 | 0.417 | 0.065 |
| Ag, BS(Ag) | -844844.596262 | -0.033 | 0.001 | 0.004 | 0.006 | -0.015 | -0.004 | 0.000 | -0.417 | -0.064 |
| Ag, BS(M1) | -844844.596332 | 0.033 | -0.001 | -0.004 | -0.006 | 0.015 | 0.004 | 0.000 | 0.418 | 0.064 |
| Ag, BS(M2) | -844844.596193 | 0.033 | -0.001 | -0.005 | -0.005 | 0.015 | 0.004 | 0.000 | 0.416 | 0.065 |

| State | Energy / eV | Ho s | Ho $p_z$ | Ho $p_x$ | Ho $p_y$ | Ho $d(z^2)$ | Ho d(xz) | Ho d(yz) | Ho $d(x^2-y^2)$ | Ho d(xy) |
|---|---|---|---|---|---|---|---|---|---|---|
| Ag, HS | -870934.360557 | 0.033 | -0.001 | -0.004 | -0.005 | 0.015 | 0.004 | 0.000 | 0.418 | 0.063 |
| Ag, BS(Ag) | -870934.359151 | 0.033 | -0.001 | -0.004 | -0.005 | 0.015 | 0.004 | 0.000 | 0.416 | 0.065 |
| Ag, BS(M1) | -870934.360126 | -0.033 | 0.001 | 0.004 | 0.005 | -0.015 | -0.004 | 0.000 | -0.418 | -0.064 |
| Ag, BS(M2) | -870934.360136 | 0.033 | -0.001 | -0.004 | -0.005 | 0.015 | 0.004 | 0.000 | 0.419 | 0.064 |

| State | Energy / eV | Er s | Er $p_z$ | Er $p_x$ | Er $p_y$ | Er $d(z^2)$ | Er d(xz) | Er d(yz) | Er $d(x^2-y^2)$ | Er d(xy) |
|---|---|---|---|---|---|---|---|---|---|---|
| Ag, HS | -897673.739213 | 0.034 | -0.001 | -0.004 | -0.005 | 0.016 | 0.004 | 0.000 | 0.417 | 0.064 |
| Ag, BS(Ag) | -897673.742544 | -0.034 | 0.001 | 0.004 | 0.006 | -0.016 | -0.004 | 0.000 | -0.417 | -0.064 |
| Ag, BS(M1) | -897673.741049 | 0.034 | -0.001 | -0.004 | -0.005 | 0.016 | 0.004 | 0.000 | 0.417 | 0.063 |
| Ag, BS(M2) | -897673.741028 | 0.034 | -0.001 | -0.004 | -0.005 | 0.016 | 0.004 | 0.000 | 0.416 | 0.064 |

| State | Energy / eV | Tm s | Tm $p_z$ | Tm $p_x$ | Tm $p_y$ | Tm $d(z^2)$ | Tm d(xz) | Tm d(yz) | Tm $d(x^2-y^2)$ | Tm d(xy) |
|---|---|---|---|---|---|---|---|---|---|---|
| Ag, HS | -925098.065351 | 0.035 | -0.001 | -0.004 | -0.005 | 0.016 | 0.003 | 0.000 | 0.415 | 0.064 |
| Ag, BS(Ag) | -925098.059992 | -0.036 | 0.001 | 0.004 | 0.006 | -0.016 | -0.003 | 0.000 | -0.416 | -0.063 |
| Ag, BS(M1) | -925098.062883 | 0.035 | -0.001 | -0.004 | -0.005 | 0.016 | 0.003 | 0.000 | 0.415 | 0.063 |
| Ag, BS(M2) | -925098.062901 | 0.035 | -0.001 | -0.004 | -0.005 | 0.016 | 0.003 | 0.000 | 0.415 | 0.064 |

| State | Energy / eV | Yb s | Yb $p_z$ | Yb $p_x$ | Yb $p_y$ | Yb $d(z^2)$ | Yb d(xz) | Yb d(yz) | Yb $d(x^2-y^2)$ | Yb d(xy) |
|---|---|---|---|---|---|---|---|---|---|---|
| Ag, HS | -953196.552443 | 0.035 | -0.001 | -0.004 | -0.005 | 0.016 | 0.004 | 0.000 | 0.416 | 0.063 |
| Ag, BS(Ag) | -953196.550022 | -0.035 | 0.001 | 0.005 | 0.006 | -0.016 | -0.004 | 0.000 | -0.416 | -0.061 |
| Ag, BS(M1) | -953196.551208 | 0.035 | -0.001 | -0.004 | -0.005 | 0.016 | 0.004 | 0.000 | 0.416 | 0.062 |
| Ag, BS(M2) | -953196.551217 | 0.035 | -0.001 | -0.004 | -0.005 | 0.016 | 0.004 | 0.000 | 0.416 | 0.062 |

## 4. Chloride $[AgM_2F_7Cl]$ clusters

**Table S9.** Total energy and Löwdin spin populations of each orbital (specified with $m_l$ number) in 3*d* or 4*f* subshell for each 3d TM or Ln cation in the clusters of $[AgM_2F_7Cl]$ type.

| State | Energy / eV | Co1 | | | | | Co2 | | | | |
|---|---|---|---|---|---|---|---|---|---|---|---|
| | | $d(z^2)$ | d(xz) | d(yz) | $d(x^2–y^2)$ | d(xy) | $d(z^2)$ | d(xz) | d(yz) | $d(x^2–y^2)$ | d(xy) |
| Cd, BS | -261795.059002 | -0.65 | -0.27 | -0.53 | -0.88 | -0.37 | 0.69 | 0.23 | 0.50 | 0.88 | 0.40 |
| Cd, HS | -261795.041490 | 0.65 | 0.28 | 0.53 | 0.88 | 0.37 | 0.69 | 0.23 | 0.50 | 0.88 | 0.39 |
| Ag, HS | -254179.004796 | 0.66 | 0.28 | 0.53 | 0.88 | 0.37 | 0.70 | 0.23 | 0.50 | 0.88 | 0.39 |
| Ag, BS(Ag) | -254179.001954 | 0.66 | 0.27 | 0.53 | 0.88 | 0.36 | 0.70 | 0.22 | 0.51 | 0.88 | 0.39 |
| Ag, BS(M1) | -254179.022114 | -0.66 | -0.26 | -0.53 | -0.88 | -0.36 | 0.70 | 0.23 | 0.50 | 0.88 | 0.39 |
| Ag, BS(M2) | -254179.020566 | 0.66 | 0.28 | 0.53 | 0.88 | 0.37 | -0.70 | -0.22 | -0.51 | -0.88 | -0.39 |

| State | Energy / eV | Ni1 | | | | | Ni2 | | | | |
|---|---|---|---|---|---|---|---|---|---|---|---|
| | | $d(z^2)$ | d(xz) | d(yz) | $d(x^2–y^2)$ | d(xy) | $d(z^2)$ | d(xz) | d(yz) | $d(x^2–y^2)$ | d(xy) |
| Cd, BS | -268777.332263 | -0.07 | -0.12 | -0.79 | -0.56 | -0.21 | 0.07 | 0.10 | 0.79 | 0.58 | 0.21 |
| Cd, HS | -268777.331983 | 0.07 | 0.12 | 0.79 | 0.56 | 0.21 | 0.07 | 0.10 | 0.79 | 0.58 | 0.21 |
| Ag, HS | -261161.271996 | 0.06 | 0.13 | 0.80 | 0.56 | 0.21 | 0.07 | 0.11 | 0.79 | 0.57 | 0.21 |
| Ag, BS(Ag) | -261161.292511 | 0.06 | 0.12 | 0.79 | 0.56 | 0.21 | 0.07 | 0.10 | 0.79 | 0.58 | 0.21 |
| Ag, BS(M1) | -261161.282920 | -0.06 | -0.12 | -0.79 | -0.56 | -0.21 | 0.07 | 0.11 | 0.79 | 0.57 | 0.21 |
| Ag, BS(M2) | -261161.282898 | 0.06 | 0.13 | 0.80 | 0.56 | 0.21 | -0.07 | -0.10 | -0.79 | -0.58 | -0.21 |

| State | Energy / eV | Pr1 | | | | | | | Pr2 | | | | | | |
|---|---|---|---|---|---|---|---|---|---|---|---|---|---|---|---|
| | | f0 | f+1 | f-1 | f+2 | f-2 | f+3 | f-3 | f0 | f+1 | f-1 | f+2 | f-2 | f+3 | f-3 |
| Cd, BS | -697872.6545164 | -0.03 | -0.51 | -0.57 | -0.02 | -0.04 | -0.17 | -0.65 | 0.04 | 0.25 | 0.82 | 0.03 | 0.06 | 0.30 | 0.52 |
| Cd, HS | -697872.6544338 | 0.03 | 0.51 | 0.57 | 0.02 | 0.04 | 0.18 | 0.64 | 0.04 | 0.25 | 0.82 | 0.03 | 0.06 | 0.30 | 0.52 |
| Ag, HS | -690256.6194200 | 0.03 | 0.51 | 0.57 | 0.02 | 0.04 | 0.19 | 0.64 | 0.04 | 0.24 | 0.82 | 0.03 | 0.06 | 0.29 | 0.52 |
| Ag, BS(Ag) | -690256.6271823 | 0.03 | 0.51 | 0.57 | 0.02 | 0.04 | 0.19 | 0.64 | 0.04 | 0.24 | 0.82 | 0.02 | 0.06 | 0.29 | 0.53 |
| Ag, BS(M1) | -690256.6233938 | -0.03 | -0.51 | -0.57 | -0.02 | -0.04 | -0.19 | -0.64 | 0.04 | 0.24 | 0.82 | 0.03 | 0.06 | 0.29 | 0.52 |
| Ag, BS(M2) | -690256.6234035 | 0.03 | 0.51 | 0.57 | 0.02 | 0.04 | 0.18 | 0.64 | -0.04 | -0.24 | -0.82 | -0.02 | -0.06 | -0.29 | -0.53 |

| State | Energy / eV | Nd1 | | | | | | | Nd2 | | | | | | |
|---|---|---|---|---|---|---|---|---|---|---|---|---|---|---|---|
| | | f0 | f+1 | f-1 | f+2 | f-2 | f+3 | f-3 | f0 | f+1 | f-1 | f+2 | f-2 | f+3 | f-3 |
| Cd, BS | -719654.7933352 | -0.22 | -0.46 | -0.61 | -0.06 | -0.72 | -0.25 | -0.67 | 0.28 | 0.70 | 0.27 | 0.59 | 0.27 | 0.41 | 0.46 |
| Cd, HS | -719654.7932265 | 0.22 | 0.46 | 0.61 | 0.06 | 0.72 | 0.25 | 0.67 | 0.28 | 0.70 | 0.27 | 0.59 | 0.27 | 0.40 | 0.46 |
| Ag, HS | -712038.7522503 | 0.22 | 0.46 | 0.61 | 0.05 | 0.72 | 0.27 | 0.65 | 0.28 | 0.71 | 0.27 | 0.58 | 0.27 | 0.41 | 0.46 |
| Ag, BS(Ag) | -712038.7539078 | 0.22 | 0.46 | 0.60 | 0.05 | 0.72 | 0.29 | 0.63 | 0.27 | 0.72 | 0.27 | 0.58 | 0.27 | 0.40 | 0.47 |
| Ag, BS(M1) | -712038.7546793 | -0.22 | -0.47 | -0.60 | -0.05 | -0.72 | -0.29 | -0.63 | 0.28 | 0.71 | 0.27 | 0.58 | 0.27 | 0.41 | 0.46 |
| Ag, BS(M2) | -712038.7516820 | 0.22 | 0.46 | 0.61 | 0.05 | 0.72 | 0.27 | 0.65 | -0.27 | -0.71 | -0.27 | -0.58 | -0.27 | -0.40 | -0.47 |

| State | Energy / eV | Pm1 | | | | | | | Pm2 | | | | | | |
|---|---|---|---|---|---|---|---|---|---|---|---|---|---|---|---|
| | | f0 | f+1 | f-1 | f+2 | f-2 | f+3 | f-3 | f0 | f+1 | f-1 | f+2 | f-2 | f+3 | f-3 |
| Cd, BS | -742023.1463532 | -0.69 | -0.48 | -0.46 | -0.95 | -0.27 | -0.51 | -0.62 | 0.67 | 0.25 | 0.71 | 0.56 | 0.65 | 0.29 | 0.84 |
| Cd, HS | -742023.1460812 | 0.69 | 0.48 | 0.46 | 0.95 | 0.27 | 0.51 | 0.62 | 0.67 | 0.25 | 0.71 | 0.56 | 0.65 | 0.29 | 0.84 |
| Ag, HS | -734407.1112579 | 0.69 | 0.48 | 0.46 | 0.95 | 0.27 | 0.51 | 0.62 | 0.67 | 0.25 | 0.71 | 0.56 | 0.65 | 0.29 | 0.84 |
| Ag, BS(Ag) | -734407.1122200 | 0.69 | 0.47 | 0.46 | 0.95 | 0.27 | 0.51 | 0.62 | 0.67 | 0.25 | 0.71 | 0.56 | 0.65 | 0.29 | 0.84 |
| Ag, BS(M1) | -734407.1120134 | -0.69 | -0.48 | -0.46 | -0.95 | -0.27 | -0.52 | -0.61 | 0.67 | 0.25 | 0.71 | 0.56 | 0.64 | 0.29 | 0.84 |
| Ag, BS(M2) | -734407.1120158 | 0.69 | 0.48 | 0.45 | 0.95 | 0.27 | 0.52 | 0.61 | -0.67 | -0.25 | -0.71 | -0.56 | -0.65 | -0.29 | -0.84 |

| State | Energy / eV | Sm1 | | | | | | | Sm2 | | | | | | |
|---|---|---|---|---|---|---|---|---|---|---|---|---|---|---|---|
| | | f0 | f+1 | f-1 | f+2 | f-2 | f+3 | f-3 | f0 | f+1 | f-1 | f+2 | f-2 | f+3 | f-3 |
| Cd, BS | -764986.8688465 | -0.89 | -0.53 | -0.68 | -0.88 | -0.47 | -0.99 | -0.53 | 0.94 | 0.40 | 0.33 | 0.97 | 0.95 | 0.83 | 0.55 |
| Cd, HS | -764986.8686585 | 0.90 | 0.54 | 0.68 | 0.88 | 0.44 | 0.99 | 0.54 | 0.94 | 0.40 | 0.33 | 0.97 | 0.95 | 0.83 | 0.54 |
| Ag, HS | -757370.8307209 | 0.90 | 0.54 | 0.68 | 0.88 | 0.44 | 0.99 | 0.54 | 0.94 | 0.40 | 0.33 | 0.97 | 0.95 | 0.83 | 0.55 |
| Ag, BS(Ag) | -757370.8294215 | 0.90 | 0.53 | 0.69 | 0.87 | 0.47 | 0.99 | 0.53 | 0.94 | 0.40 | 0.33 | 0.97 | 0.95 | 0.83 | 0.55 |
| Ag, BS(M1) | -757370.8303880 | -0.90 | -0.54 | -0.69 | -0.88 | -0.43 | -0.99 | -0.56 | 0.94 | 0.40 | 0.33 | 0.97 | 0.95 | 0.83 | 0.55 |
| Ag, BS(M2) | -757370.8300333 | 0.90 | 0.53 | 0.68 | 0.87 | 0.48 | 0.99 | 0.53 | -0.94 | -0.40 | -0.33 | -0.97 | -0.95 | -0.83 | -0.55 |

| State | Energy / eV | Eu1 | | | | | | | Eu2 | | | | | | |
|---|---|---|---|---|---|---|---|---|---|---|---|---|---|---|---|
| | | f0 | f+1 | f-1 | f+2 | f-2 | f+3 | f-3 | f0 | f+1 | f-1 | f+2 | f-2 | f+3 | f-3 |
| Cd, BS | -788557.6839468 | -0.94 | -0.61 | -0.70 | -0.98 | -0.99 | -0.98 | -0.78 | 0.94 | 0.39 | 0.92 | 0.98 | 0.99 | 0.85 | 0.91 |
| Cd, HS | -788557.6837695 | 0.94 | 0.61 | 0.70 | 0.98 | 0.99 | 0.98 | 0.78 | 0.94 | 0.39 | 0.92 | 0.98 | 0.99 | 0.85 | 0.91 |
| Ag, HS | -780941.6451370 | 0.94 | 0.61 | 0.70 | 0.98 | 0.99 | 0.98 | 0.78 | 0.94 | 0.39 | 0.92 | 0.98 | 0.99 | 0.86 | 0.91 |
| Ag, BS(Ag) | -780941.6429831 | 0.94 | 0.60 | 0.71 | 0.98 | 0.99 | 0.99 | 0.78 | 0.94 | 0.39 | 0.92 | 0.98 | 0.99 | 0.85 | 0.91 |
| Ag, BS(M1) | -780941.6442593 | -0.94 | -0.60 | -0.71 | -0.98 | -0.99 | -0.99 | -0.78 | 0.94 | 0.39 | 0.92 | 0.98 | 0.99 | 0.86 | 0.91 |
| Ag, BS(M2) | -780941.6442483 | 0.94 | 0.61 | 0.70 | 0.98 | 0.99 | 0.98 | 0.78 | -0.94 | -0.39 | -0.92 | -0.98 | -0.99 | -0.85 | -0.92 |

| State | Energy / eV | Gd1 | | | | | | | Gd2 | | | | | | |
|---|---|---|---|---|---|---|---|---|---|---|---|---|---|---|---|
| | | f0 | f+1 | f-1 | f+2 | f-2 | f+3 | f-3 | f0 | f+1 | f-1 | f+2 | f-2 | f+3 | f-3 |
| Cd, BS | -812749.0477548 | -0.99 | -0.99 | -0.99 | -0.99 | -0.99 | -0.99 | -0.99 | 0.99 | 0.98 | 0.99 | 0.99 | 0.99 | 0.99 | 0.99 |
| Cd, HS | -812749.0474756 | 0.99 | 0.99 | 0.99 | 0.99 | 0.99 | 0.99 | 0.99 | 0.99 | 0.98 | 0.99 | 0.99 | 0.99 | 0.99 | 0.99 |
| Ag, HS | -805132.9916182 | 0.99 | 0.99 | 0.99 | 0.99 | 0.99 | 0.99 | 0.99 | 0.99 | 0.98 | 0.99 | 0.99 | 0.99 | 0.99 | 0.99 |
| Ag, BS(Ag) | -805132.9897960 | 0.99 | 0.99 | 0.99 | 0.99 | 0.99 | 0.99 | 0.99 | 0.99 | 0.98 | 0.99 | 0.99 | 0.99 | 0.99 | 0.99 |
| Ag, BS(M1) | -805132.9909971 | -0.99 | -0.99 | -0.99 | -0.99 | -0.99 | -0.99 | -0.99 | 0.99 | 0.98 | 0.99 | 0.99 | 0.99 | 0.99 | 0.99 |
| Ag, BS(M2) | -805132.9909952 | 0.99 | 0.99 | 0.99 | 0.99 | 0.99 | 0.99 | 0.99 | -0.99 | -0.98 | -0.99 | -0.99 | -0.99 | -0.99 | -0.99 |

| State | Energy / eV | Tb1 | | | | | | | Tb2 | | | | | | |
|---|---|---|---|---|---|---|---|---|---|---|---|---|---|---|---|
| | | f0 | f+1 | f-1 | f+2 | f-2 | f+3 | f-3 | f0 | f+1 | f-1 | f+2 | f-2 | f+3 | f-3 |
| Cd, BS | -837549.2105632 | -0.98 | -0.93 | -0.52 | -0.98 | -0.95 | -0.81 | -0.75 | 0.95 | 0.72 | 0.70 | 0.98 | 0.98 | 0.81 | 0.77 |
| Cd, HS | -837549.2103629 | 0.98 | 0.93 | 0.52 | 0.98 | 0.95 | 0.81 | 0.75 | 0.95 | 0.72 | 0.70 | 0.98 | 0.98 | 0.81 | 0.77 |
| Ag, HS | -829933.1530550 | 0.98 | 0.93 | 0.52 | 0.97 | 0.95 | 0.80 | 0.76 | 0.96 | 0.72 | 0.71 | 0.97 | 0.99 | 0.84 | 0.75 |
| Ag, BS(Ag) | -829933.1494068 | 0.98 | 0.93 | 0.52 | 0.97 | 0.95 | 0.80 | 0.76 | 0.96 | 0.72 | 0.70 | 0.97 | 0.99 | 0.82 | 0.76 |
| Ag, BS(M1) | -829933.1524994 | -0.98 | -0.93 | -0.52 | -0.98 | -0.95 | -0.80 | -0.76 | 0.96 | 0.72 | 0.71 | 0.97 | 0.99 | 0.83 | 0.75 |
| Ag, BS(M2) | -829933.1503849 | 0.98 | 0.93 | 0.53 | 0.97 | 0.95 | 0.80 | 0.76 | -0.96 | -0.72 | -0.70 | -0.97 | -0.99 | -0.82 | -0.76 |

| State | Energy / eV | Dy1 | | | | | | | Dy2 | | | | | | |
|---|---|---|---|---|---|---|---|---|---|---|---|---|---|---|---|
| | | f0 | f+1 | f-1 | f+2 | f-2 | f+3 | f-3 | f0 | f+1 | f-1 | f+2 | f-2 | f+3 | f-3 |
| Cd, BS | -862990.7620693 | -0.94 | -0.44 | -0.44 | -0.97 | -0.94 | -0.69 | -0.51 | 0.95 | 0.67 | 0.19 | 0.97 | 0.94 | 0.82 | 0.37 |
| Cd, HS | -862990.7619604 | 0.94 | 0.44 | 0.44 | 0.97 | 0.94 | 0.69 | 0.51 | 0.95 | 0.67 | 0.19 | 0.97 | 0.94 | 0.82 | 0.37 |
| Ag, HS | -855374.6983829 | 0.94 | 0.44 | 0.44 | 0.97 | 0.94 | 0.68 | 0.52 | 0.95 | 0.68 | 0.18 | 0.97 | 0.94 | 0.83 | 0.37 |
| Ag, BS(Ag) | -855374.6941362 | 0.94 | 0.44 | 0.44 | 0.97 | 0.94 | 0.68 | 0.52 | 0.95 | 0.68 | 0.18 | 0.97 | 0.94 | 0.83 | 0.37 |
| Ag, BS(M1) | -855374.6963800 | -0.94 | -0.44 | -0.44 | -0.97 | -0.94 | -0.68 | -0.52 | 0.95 | 0.68 | 0.18 | 0.97 | 0.94 | 0.83 | 0.37 |
| Ag, BS(M2) | -855374.6963801 | 0.94 | 0.44 | 0.44 | 0.97 | 0.94 | 0.68 | 0.52 | -0.96 | -0.67 | -0.18 | -0.97 | -0.94 | -0.83 | -0.37 |

| State | Energy / eV | Ho1 | | | | | | | Ho2 | | | | | | |
|---|---|---|---|---|---|---|---|---|---|---|---|---|---|---|---|
| | | f0 | f+1 | f-1 | f+2 | f-2 | f+3 | f-3 | f0 | f+1 | f-1 | f+2 | f-2 | f+3 | f-3 |
| Cd, BS | -889080.4625254 | -0.67 | -0.16 | -0.85 | -0.28 | -0.86 | -0.78 | -0.34 | 0.59 | 0.43 | 0.70 | 0.29 | 0.81 | 0.63 | 0.49 |
| Cd, HS | -889080.4624879 | 0.67 | 0.16 | 0.85 | 0.28 | 0.86 | 0.78 | 0.33 | 0.59 | 0.43 | 0.70 | 0.29 | 0.81 | 0.63 | 0.49 |
| Ag, HS | -881464.3917648 | 0.68 | 0.15 | 0.85 | 0.28 | 0.86 | 0.78 | 0.34 | 0.59 | 0.43 | 0.70 | 0.28 | 0.82 | 0.64 | 0.48 |
| Ag, BS(Ag) | -881464.3908812 | 0.67 | 0.16 | 0.85 | 0.28 | 0.87 | 0.78 | 0.33 | 0.59 | 0.44 | 0.70 | 0.29 | 0.81 | 0.63 | 0.49 |
| Ag, BS(M1) | -881464.3913977 | -0.68 | -0.16 | -0.85 | -0.28 | -0.87 | -0.78 | -0.33 | 0.59 | 0.43 | 0.70 | 0.28 | 0.82 | 0.64 | 0.48 |
| Ag, BS(M2) | -881464.3913346 | 0.67 | 0.16 | 0.85 | 0.28 | 0.86 | 0.78 | 0.34 | -0.59 | -0.44 | -0.69 | -0.29 | -0.81 | -0.63 | -0.49 |

**Table S10.** Total energy and Löwdin spin populations of each orbital (specified with $m_l$ number) in 4*s*, 4*p* and 3*d* subshells for the silver cation in the clusters of $[AgM_2F_7Cl]$ type.

| State | Energy / eV | Co | | | | | | | | |
|---|---|---|---|---|---|---|---|---|---|---|
| | | s | $p_z$ | $p_x$ | $p_y$ | $d(z^2)$ | d(xz) | d(yz) | $d(x^2–y^2)$ | d(xy) |
| Ag, HS | -254179.004796 | -0.016 | -0.005 | -0.003 | -0.003 | 0.364 | 0.000 | 0.013 | 0.148 | 0.000 |
| Ag, BS(Ag) | -254179.001954 | 0.018 | 0.007 | 0.004 | 0.001 | -0.367 | 0.000 | -0.014 | -0.150 | 0.000 |
| Ag, BS(M1) | -254179.022114 | -0.017 | -0.006 | -0.003 | -0.002 | 0.365 | 0.000 | 0.014 | 0.149 | 0.000 |
| Ag, BS(M2) | -254179.020566 | -0.017 | -0.006 | -0.003 | -0.002 | 0.366 | 0.000 | 0.014 | 0.149 | 0.000 |

| State | Energy / eV | Ni | | | | | | | | |
|---|---|---|---|---|---|---|---|---|---|---|
| | | s | $p_z$ | $p_x$ | $p_y$ | $d(z^2)$ | d(xz) | d(yz) | $d(x^2–y^2)$ | d(xy) |
| Ag, HS | -261161.271996 | -0.005 | -0.005 | -0.005 | -0.002 | 0.381 | 0.001 | 0.000 | 0.113 | 0.039 |
| Ag, BS(Ag) | -261161.292511 | 0.010 | 0.007 | 0.005 | 0.001 | -0.380 | 0.000 | 0.000 | -0.112 | -0.038 |
| Ag, BS(M1) | -261161.282920 | -0.008 | -0.006 | -0.005 | -0.002 | 0.380 | 0.001 | 0.000 | 0.112 | 0.038 |
| Ag, BS(M2) | -261161.282898 | -0.008 | -0.006 | -0.005 | -0.002 | 0.380 | 0.000 | 0.000 | 0.112 | 0.038 |

| State | Energy / eV | Pr | | | | | | | | |
|---|---|---|---|---|---|---|---|---|---|---|
| | | s | $p_z$ | $p_x$ | $p_y$ | $d(z^2)$ | d(xz) | d(yz) | $d(x^2–y^2)$ | d(xy) |
| Ag, HS | -690256.6194200 | -0.008 | -0.002 | -0.004 | 0.008 | 0.010 | 0.004 | 0.000 | 0.409 | 0.064 |
| Ag, BS(Ag) | -690256.6271823 | 0.009 | 0.002 | 0.003 | -0.008 | -0.010 | -0.004 | 0.000 | -0.405 | -0.063 |
| Ag, BS(M1) | -690256.6233938 | -0.008 | -0.002 | -0.003 | 0.008 | 0.010 | 0.004 | 0.000 | 0.407 | 0.064 |
| Ag, BS(M2) | -690256.6234035 | -0.008 | -0.002 | -0.003 | 0.008 | 0.010 | 0.004 | 0.000 | 0.407 | 0.064 |

| State | Energy / eV | Nd | | | | | | | | |
|---|---|---|---|---|---|---|---|---|---|---|
| | | s | $p_z$ | $p_x$ | $p_y$ | $d(z^2)$ | d(xz) | d(yz) | $d(x^2–y^2)$ | d(xy) |
| Ag, HS | -712038.7522503 | -0.008 | -0.002 | -0.003 | 0.008 | 0.010 | 0.004 | 0.000 | 0.408 | 0.066 |
| Ag, BS(Ag) | -712038.7539078 | 0.008 | 0.002 | 0.003 | -0.008 | -0.010 | -0.004 | 0.000 | -0.408 | -0.066 |
| Ag, BS(M1) | -712038.7546793 | -0.008 | -0.002 | -0.003 | 0.008 | 0.010 | 0.004 | 0.000 | 0.408 | 0.066 |
| Ag, BS(M2) | -712038.7516820 | -0.008 | -0.002 | -0.003 | 0.008 | 0.010 | 0.004 | 0.000 | 0.408 | 0.066 |

| State | Energy / eV | Pm | | | | | | | | |
|---|---|---|---|---|---|---|---|---|---|---|
| | | s | $p_z$ | $p_x$ | $p_y$ | $d(z^2)$ | d(xz) | d(yz) | $d(x^2–y^2)$ | d(xy) |
| Ag, HS | -734407.1112579 | -0.008 | -0.002 | -0.004 | 0.008 | 0.010 | 0.004 | 0.000 | 0.408 | 0.064 |
| Ag, BS(Ag) | -734407.1122200 | 0.009 | 0.002 | 0.003 | -0.008 | -0.010 | -0.004 | 0.000 | -0.407 | -0.064 |
| Ag, BS(M1) | -734407.1120134 | -0.009 | -0.002 | -0.003 | 0.008 | 0.010 | 0.004 | 0.000 | 0.407 | 0.064 |
| Ag, BS(M2) | -734407.1120158 | -0.009 | -0.002 | -0.003 | 0.008 | 0.010 | 0.004 | 0.000 | 0.408 | 0.064 |

| State | Energy / eV | Sm | | | | | | | | |
|---|---|---|---|---|---|---|---|---|---|---|
| | | s | $p_z$ | $p_x$ | $p_y$ | $d(z^2)$ | d(xz) | d(yz) | $d(x^2–y^2)$ | d(xy) |
| Ag, HS | -757370.8307209 | -0.009 | -0.002 | -0.004 | 0.008 | 0.010 | 0.004 | 0.001 | 0.398 | 0.072 |
| Ag, BS(Ag) | -757370.8294215 | 0.009 | 0.002 | 0.003 | -0.009 | -0.010 | -0.004 | 0.000 | -0.398 | -0.072 |
| Ag, BS(M1) | -757370.8303880 | -0.009 | -0.002 | -0.003 | 0.009 | 0.010 | 0.004 | 0.001 | 0.398 | 0.072 |
| Ag, BS(M2) | -757370.8300333 | -0.009 | -0.002 | -0.003 | 0.009 | 0.010 | 0.004 | 0.000 | 0.398 | 0.072 |

| State | Energy / eV | Eu | | | | | | | | |
|---|---|---|---|---|---|---|---|---|---|---|
| | | s | $p_z$ | $p_x$ | $p_y$ | $d(z^2)$ | d(xz) | d(yz) | $d(x^2–y^2)$ | d(xy) |
| Ag, HS | -780941.6451370 | -0.009 | -0.002 | -0.004 | 0.009 | 0.010 | 0.004 | 0.000 | 0.405 | 0.063 |
| Ag, BS(Ag) | -780941.6429831 | 0.009 | 0.002 | 0.003 | -0.009 | -0.010 | -0.004 | 0.000 | -0.405 | -0.063 |
| Ag, BS(M1) | -780941.6442593 | -0.009 | -0.002 | -0.003 | 0.009 | 0.010 | 0.004 | 0.000 | 0.405 | 0.063 |
| Ag, BS(M2) | -780941.6442483 | -0.009 | -0.002 | -0.003 | 0.009 | 0.010 | 0.004 | 0.000 | 0.405 | 0.063 |

| State | Energy / eV | Gd | | | | | | | | |
|---|---|---|---|---|---|---|---|---|---|---|
| | | s | $p_z$ | $p_x$ | $p_y$ | $d(z^2)$ | d(xz) | d(yz) | $d(x^2–y^2)$ | d(xy) |
| Ag, HS | -805132.9916182 | -0.010 | -0.002 | -0.003 | 0.011 | 0.010 | 0.004 | 0.000 | 0.405 | 0.062 |
| Ag, BS(Ag) | -805132.9897960 | 0.010 | 0.002 | 0.003 | -0.010 | -0.010 | -0.004 | 0.000 | -0.405 | -0.062 |
| Ag, BS(M1) | -805132.9909971 | -0.010 | -0.002 | -0.003 | 0.010 | 0.010 | 0.004 | 0.000 | 0.405 | 0.062 |
| Ag, BS(M2) | -805132.9909952 | -0.010 | -0.002 | -0.003 | 0.011 | 0.010 | 0.004 | 0.000 | 0.405 | 0.062 |

| State | Energy / eV | Tb | | | | | | | | |
|---|---|---|---|---|---|---|---|---|---|---|
| | | s | $p_z$ | $p_x$ | $p_y$ | $d(z^2)$ | d(xz) | d(yz) | $d(x^2–y^2)$ | d(xy) |
| Ag, HS | -829933.1530550 | -0.009 | -0.002 | -0.003 | 0.010 | 0.011 | 0.004 | 0.000 | 0.393 | 0.072 |
| Ag, BS(Ag) | -829933.1494068 | 0.009 | 0.002 | 0.003 | -0.010 | -0.011 | -0.004 | 0.000 | -0.394 | -0.072 |
| Ag, BS(M1) | -829933.1524994 | -0.009 | -0.002 | -0.003 | 0.010 | 0.010 | 0.004 | 0.000 | 0.393 | 0.072 |
| Ag, BS(M2) | -829933.1503849 | -0.009 | -0.002 | -0.003 | 0.010 | 0.011 | 0.004 | 0.000 | 0.394 | 0.072 |

| State | Energy / eV | Dy | | | | | | | | |
|---|---|---|---|---|---|---|---|---|---|---|
| | | s | $p_z$ | $p_x$ | $p_y$ | $d(z^2)$ | d(xz) | d(yz) | $d(x^2–y^2)$ | d(xy) |
| Ag, HS | -855374.6983829 | -0.009 | -0.002 | -0.003 | 0.011 | 0.011 | 0.004 | 0.000 | 0.403 | 0.062 |
| Ag, BS(Ag) | -855374.6941362 | 0.009 | 0.002 | 0.003 | -0.011 | -0.011 | -0.004 | 0.000 | -0.403 | -0.062 |
| Ag, BS(M1) | -855374.6963800 | -0.009 | -0.002 | -0.003 | 0.011 | 0.011 | 0.004 | 0.000 | 0.403 | 0.062 |
| Ag, BS(M2) | -855374.6963801 | -0.009 | -0.002 | -0.003 | 0.011 | 0.011 | 0.004 | 0.000 | 0.403 | 0.062 |

| State | Energy / eV | Ho | | | | | | | | |
|---|---|---|---|---|---|---|---|---|---|---|
| | | s | $p_z$ | $p_x$ | $p_y$ | $d(z^2)$ | d(xz) | d(yz) | $d(x^2–y^2)$ | d(xy) |
| Ag, HS | -881464.3917648 | -0.009 | -0.002 | -0.003 | 0.011 | 0.011 | 0.004 | 0.000 | 0.402 | 0.062 |
| Ag, BS(Ag) | -881464.3908812 | 0.008 | 0.002 | 0.003 | -0.011 | -0.011 | -0.004 | 0.000 | -0.402 | -0.062 |
| Ag, BS(M1) | -881464.3913977 | -0.008 | -0.002 | -0.003 | 0.011 | 0.011 | 0.004 | 0.000 | 0.402 | 0.062 |
| Ag, BS(M2) | -881464.3913346 | -0.008 | -0.002 | -0.003 | 0.011 | 0.011 | 0.004 | 0.000 | 0.402 | 0.062 |

## 5. Fluoride [$AgM_2F_8$] clusters

**Table S11.** Total energy and Löwdin spin populations of each orbital (specified with $m_l$ number) in 3*d* or 4*f* subshell for each 3d TM or Ln cation in the clusters of [$AgM_2F_8$] type.

| State | Energy / eV | Co1 | | | | | Co2 | | | | |
|---|---|---|---|---|---|---|---|---|---|---|---|
| | | d(z²) | d(xz) | d(yz) | d(x²–y²) | d(xy) | d(z²) | d(xz) | d(yz) | d(x²–y²) | d(xy) |
| Cd, BS | -251932.535205 | -0.88 | -0.08 | -0.45 | -0.85 | -0.47 | 0.88 | 0.08 | 0.42 | 0.83 | 0.50 |
| Cd, HS | -251932.525115 | 0.88 | 0.08 | 0.45 | 0.85 | 0.47 | 0.88 | 0.08 | 0.42 | 0.83 | 0.50 |
| Ag, HS | -244316.366165 | 0.87 | 0.10 | 0.46 | 0.86 | 0.46 | 0.88 | 0.10 | 0.43 | 0.84 | 0.49 |
| Ag, BS(Ag) | -244316.350967 | 0.88 | 0.08 | 0.45 | 0.85 | 0.46 | 0.88 | 0.08 | 0.43 | 0.83 | 0.50 |
| Ag, BS(M1) | -244316.371471 | -0.88 | -0.08 | -0.45 | -0.85 | -0.47 | 0.88 | 0.10 | 0.43 | 0.84 | 0.49 |
| Ag, BS(M2) | -244316.364831 | 0.87 | 0.10 | 0.45 | 0.86 | 0.46 | -0.88 | -0.08 | -0.42 | -0.83 | -0.50 |

| State | Energy / eV | Ni1 | | | | | Ni2 | | | | |
|---|---|---|---|---|---|---|---|---|---|---|---|
| | | d(z²) | d(xz) | d(yz) | d(x²–y²) | d(xy) | d(z²) | d(xz) | d(yz) | d(x²–y²) | d(xy) |
| Cd, BS | -258914.678628 | -0.42 | -0.02 | -0.44 | -0.57 | -0.32 | 0.39 | 0.10 | 0.38 | 0.68 | 0.22 |
| Cd, HS | -258914.675880 | 0.41 | 0.02 | 0.44 | 0.57 | 0.32 | 0.39 | 0.10 | 0.39 | 0.68 | 0.22 |
| Ag, HS | -251298.501134 | 0.41 | 0.02 | 0.45 | 0.57 | 0.32 | 0.38 | 0.09 | 0.40 | 0.68 | 0.21 |
| Ag, BS(Ag) | -251298.512098 | 0.41 | 0.02 | 0.45 | 0.57 | 0.31 | 0.38 | 0.10 | 0.39 | 0.68 | 0.22 |
| Ag, BS(M1) | -251298.509556 | -0.41 | -0.02 | -0.44 | -0.57 | -0.31 | 0.39 | 0.09 | 0.39 | 0.68 | 0.21 |
| Ag, BS(M2) | -251298.509841 | 0.41 | 0.02 | 0.44 | 0.57 | 0.32 | -0.38 | -0.10 | -0.39 | -0.68 | -0.22 |

| State | Energy / eV | Cu1 | | | | | Cu2 | | | | |
|---|---|---|---|---|---|---|---|---|---|---|---|
| | | d(z²) | d(xz) | d(yz) | d(x²–y²) | d(xy) | d(z²) | d(xz) | d(yz) | d(x²–y²) | d(xy) |
| Cd, BS | -266275.939785 | -0.45 | 0.00 | -0.05 | -0.11 | -0.12 | 0.46 | 0.01 | 0.04 | 0.11 | 0.12 |
| Cd, HS | -266275.934559 | 0.45 | 0.00 | 0.05 | 0.11 | 0.12 | 0.46 | 0.01 | 0.04 | 0.11 | 0.12 |
| Ag, HS | -258659.782771 | 0.45 | 0.00 | 0.05 | 0.11 | 0.12 | 0.46 | 0.01 | 0.04 | 0.11 | 0.12 |
| Ag, BS(Ag) | -258659.818013 | 0.45 | 0.00 | 0.05 | 0.11 | 0.12 | 0.45 | 0.01 | 0.04 | 0.11 | 0.12 |
| Ag, BS(M1) | -258659.807810 | -0.45 | 0.00 | -0.05 | -0.11 | -0.12 | 0.46 | 0.01 | 0.04 | 0.11 | 0.12 |
| Ag, BS(M2) | -258659.808730 | 0.45 | 0.00 | 0.05 | 0.11 | 0.12 | -0.45 | -0.01 | -0.04 | -0.11 | -0.12 |

| State | Energy / eV | Nd1 | | | | | | | Nd2 | | | | | | |
|---|---|---|---|---|---|---|---|---|---|---|---|---|---|---|---|
| | | f0 | f+1 | f-1 | f+2 | f-2 | f+3 | f-3 | f0 | f+1 | f-1 | f+2 | f-2 | f+3 | f-3 |
| Cd, BS | -709792.8836851 | -0.29 | -0.61 | -0.35 | -0.11 | -0.71 | -0.77 | -0.14 | 0.28 | 0.81 | 0.16 | 0.79 | 0.07 | 0.06 | 0.81 |
| Cd, HS | -709792.8835880 | 0.29 | 0.61 | 0.35 | 0.11 | 0.71 | 0.77 | 0.14 | 0.28 | 0.81 | 0.16 | 0.79 | 0.07 | 0.06 | 0.82 |
| Ag, HS | -702176.6803322 | 0.29 | 0.61 | 0.35 | 0.11 | 0.71 | 0.77 | 0.14 | 0.28 | 0.81 | 0.16 | 0.78 | 0.07 | 0.06 | 0.81 |
| Ag, BS(Ag) | -702176.6838464 | 0.28 | 0.62 | 0.36 | 0.11 | 0.70 | 0.78 | 0.13 | 0.28 | 0.81 | 0.16 | 0.78 | 0.07 | 0.06 | 0.82 |
| Ag, BS(M1) | -702176.6840387 | -0.28 | -0.62 | -0.36 | -0.11 | -0.70 | -0.78 | -0.13 | 0.28 | 0.81 | 0.16 | 0.78 | 0.07 | 0.06 | 0.81 |
| Ag, BS(M2) | -702176.6802646 | 0.28 | 0.62 | 0.36 | 0.11 | 0.70 | 0.78 | 0.13 | -0.25 | -0.84 | -0.18 | -0.77 | -0.06 | -0.05 | -0.83 |

| State | Energy / eV | Pm1 | | | | | | | Pm2 | | | | | | |
|---|---|---|---|---|---|---|---|---|---|---|---|---|---|---|---|
| | | f0 | f+1 | f-1 | f+2 | f-2 | f+3 | f-3 | f0 | f+1 | f-1 | f+2 | f-2 | f+3 | f-3 |
| Cd, BS | -732161.2758065 | -0.71 | -0.63 | -0.27 | -0.87 | -0.37 | -0.95 | -0.18 | 0.70 | 0.37 | 0.54 | 0.84 | 0.39 | 0.21 | 0.92 |
| Cd, HS | -732161.2755478 | 0.71 | 0.63 | 0.28 | 0.87 | 0.37 | 0.94 | 0.18 | 0.70 | 0.37 | 0.54 | 0.84 | 0.39 | 0.21 | 0.92 |
| Ag, HS | -724545.0576993 | 0.71 | 0.63 | 0.28 | 0.87 | 0.36 | 0.94 | 0.18 | 0.70 | 0.37 | 0.54 | 0.84 | 0.39 | 0.21 | 0.92 |
| Ag, BS(Ag) | -724545.0585766 | 0.71 | 0.62 | 0.28 | 0.87 | 0.36 | 0.94 | 0.19 | 0.70 | 0.37 | 0.54 | 0.84 | 0.39 | 0.20 | 0.92 |
| Ag, BS(M1) | -724545.0584061 | -0.71 | -0.62 | -0.28 | -0.88 | -0.36 | -0.93 | -0.19 | 0.70 | 0.37 | 0.54 | 0.84 | 0.38 | 0.21 | 0.92 |
| Ag, BS(M2) | -724545.0584184 | 0.71 | 0.63 | 0.28 | 0.87 | 0.37 | 0.94 | 0.18 | -0.70 | -0.37 | -0.55 | -0.84 | -0.39 | -0.20 | -0.93 |

| State | Energy / eV | Sm1 | | | | | | | Sm2 | | | | | | |
|---|---|---|---|---|---|---|---|---|---|---|---|---|---|---|---|
| | | f0 | f+1 | f-1 | f+2 | f-2 | f+3 | f-3 | f0 | f+1 | f-1 | f+2 | f-2 | f+3 | f-3 |
| Cd, BS | -755125.0453333 | -0.95 | -0.60 | -0.12 | -0.95 | -0.97 | -0.65 | -0.72 | 0.95 | 0.54 | 0.21 | 0.94 | 0.97 | 0.54 | 0.83 |
| Cd, HS | -755125.0450346 | 0.95 | 0.61 | 0.13 | 0.95 | 0.97 | 0.65 | 0.72 | 0.94 | 0.54 | 0.21 | 0.94 | 0.97 | 0.54 | 0.83 |
| Ag, HS | -747508.8080837 | 0.95 | 0.61 | 0.13 | 0.95 | 0.97 | 0.65 | 0.72 | 0.94 | 0.54 | 0.21 | 0.94 | 0.97 | 0.54 | 0.83 |
| Ag, BS(Ag) | -747508.8064226 | 0.95 | 0.61 | 0.12 | 0.95 | 0.97 | 0.64 | 0.73 | 0.95 | 0.54 | 0.21 | 0.94 | 0.97 | 0.54 | 0.83 |
| Ag, BS(M1) | -747508.8075397 | -0.95 | -0.61 | -0.12 | -0.95 | -0.97 | -0.64 | -0.73 | 0.95 | 0.54 | 0.21 | 0.94 | 0.97 | 0.54 | 0.83 |
| Ag, BS(M2) | -747508.8075367 | 0.95 | 0.61 | 0.12 | 0.95 | 0.97 | 0.64 | 0.73 | -0.95 | -0.54 | -0.21 | -0.94 | -0.97 | -0.54 | -0.83 |

**Table S12.** Total energy and Löwdin spin populations of each orbital (specified with $m_l$ number) in 4*s*, 4*p* and 3*d* subshells for the silver cation in the clusters of [$AgM_2F_8$] type.

| State | Energy / eV | s | $p_z$ | $p_x$ | $p_y$ | $d(z^2)$ | d(xz) | d(yz) | $d(x^2–y^2)$ | d(xy) |
|---|---|---|---|---|---|---|---|---|---|---|
| | | **Co** | | | | | | | | |
| Ag, HS | -244316.366165 | -0.021 | -0.005 | 0.002 | -0.003 | 0.370 | 0.001 | 0.013 | 0.159 | 0.000 |
| Ag, BS(Ag) | -244316.350967 | 0.023 | 0.007 | 0.001 | 0.001 | -0.383 | 0.000 | -0.014 | -0.168 | 0.000 |
| Ag, BS(M1) | -244316.371471 | -0.022 | -0.006 | 0.001 | -0.002 | 0.375 | 0.001 | 0.014 | 0.162 | 0.000 |
| Ag, BS(M2) | -244316.364831 | -0.022 | -0.006 | 0.001 | -0.002 | 0.378 | 0.000 | 0.014 | 0.164 | 0.000 |

| State | Energy / eV | s | $p_z$ | $p_x$ | $p_y$ | $d(z^2)$ | d(xz) | d(yz) | $d(x^2–y^2)$ | d(xy) |
|---|---|---|---|---|---|---|---|---|---|---|
| | | **Ni** | | | | | | | | |
| Ag, HS | -251298.501134 | -0.018 | -0.005 | -0.001 | -0.002 | 0.390 | 0.003 | 0.009 | 0.127 | 0.049 |
| Ag, BS(Ag) | -251298.512098 | 0.018 | 0.007 | 0.001 | 0.001 | -0.388 | -0.002 | -0.009 | -0.126 | -0.049 |
| Ag, BS(M1) | -251298.509556 | -0.018 | -0.006 | -0.001 | -0.002 | 0.389 | 0.003 | 0.009 | 0.126 | 0.050 |
| Ag, BS(M2) | -251298.509841 | -0.018 | -0.006 | -0.001 | -0.002 | 0.389 | 0.002 | 0.009 | 0.127 | 0.049 |

| State | Energy / eV | s | $p_z$ | $p_x$ | $p_y$ | $d(z^2)$ | d(xz) | d(yz) | $d(x^2–y^2)$ | d(xy) |
|---|---|---|---|---|---|---|---|---|---|---|
| | | **Cu** | | | | | | | | |
| Ag, HS | -258659.782771 | -0.019 | -0.006 | -0.001 | -0.002 | 0.401 | 0.000 | 0.003 | 0.169 | 0.000 |
| Ag, BS(Ag) | -258659.818013 | 0.017 | 0.007 | 0.003 | 0.001 | -0.396 | 0.000 | -0.003 | -0.165 | 0.000 |
| Ag, BS(M1) | -258659.807810 | -0.018 | -0.006 | -0.002 | -0.002 | 0.398 | 0.000 | 0.003 | 0.167 | 0.000 |
| Ag, BS(M2) | -258659.808730 | -0.018 | -0.006 | -0.002 | -0.002 | 0.398 | 0.000 | 0.003 | 0.167 | 0.000 |

| State | Energy / eV | s | $p_z$ | $p_x$ | $p_y$ | $d(z^2)$ | d(xz) | d(yz) | $d(x^2–y^2)$ | d(xy) |
|---|---|---|---|---|---|---|---|---|---|---|
| | | **Nd** | | | | | | | | |
| Ag, HS | -702176.6803322 | -0.016 | -0.002 | -0.003 | 0.020 | 0.012 | 0.005 | 0.000 | 0.403 | 0.066 |
| Ag, BS(Ag) | -702176.6838464 | 0.016 | 0.002 | 0.002 | -0.020 | -0.012 | -0.005 | 0.000 | -0.402 | -0.065 |
| Ag, BS(M1) | -702176.6840387 | -0.016 | -0.002 | -0.002 | 0.020 | 0.012 | 0.005 | 0.000 | 0.402 | 0.065 |
| Ag, BS(M2) | -702176.6802646 | -0.016 | -0.002 | -0.002 | 0.020 | 0.012 | 0.005 | 0.000 | 0.403 | 0.066 |

| State | Energy / eV | s | $p_z$ | $p_x$ | $p_y$ | $d(z^2)$ | d(xz) | d(yz) | $d(x^2–y^2)$ | d(xy) |
|---|---|---|---|---|---|---|---|---|---|---|
| | | **Pm** | | | | | | | | |
| Ag, HS | -724545.0576993 | -0.016 | -0.002 | -0.002 | 0.022 | 0.012 | 0.004 | 0.000 | 0.402 | 0.062 |
| Ag, BS(Ag) | -724545.0585766 | 0.017 | 0.002 | 0.002 | -0.022 | -0.012 | -0.004 | 0.000 | -0.402 | -0.062 |
| Ag, BS(M1) | -724545.0584061 | -0.016 | -0.002 | -0.002 | 0.022 | 0.012 | 0.004 | 0.000 | 0.402 | 0.062 |
| Ag, BS(M2) | -724545.0584184 | -0.016 | -0.002 | -0.002 | 0.022 | 0.012 | 0.004 | 0.000 | 0.402 | 0.062 |

| State | Energy / eV | s | $p_z$ | $p_x$ | $p_y$ | $d(z^2)$ | d(xz) | d(yz) | $d(x^2–y^2)$ | d(xy) |
|---|---|---|---|---|---|---|---|---|---|---|
| | | **Sm** | | | | | | | | |
| Ag, HS | -747508.8080837 | -0.016 | -0.002 | -0.002 | 0.022 | 0.013 | 0.004 | 0.000 | 0.402 | 0.061 |
| Ag, BS(Ag) | -747508.8064226 | 0.016 | 0.002 | 0.002 | -0.023 | -0.013 | -0.004 | 0.000 | -0.402 | -0.061 |
| Ag, BS(M1) | -747508.8075397 | -0.016 | -0.002 | -0.002 | 0.022 | 0.013 | 0.004 | 0.000 | 0.402 | 0.061 |
| Ag, BS(M2) | -747508.8075367 | -0.016 | -0.002 | -0.002 | 0.022 | 0.013 | 0.004 | 0.000 | 0.402 | 0.061 |

## 6. XYZ geometries for selected clusters

XYZ cartesian coordinates provided in angstroms for all clusters with 3d metals and the following clusters with Ln: Eu/$O^{2-}$, Gd/$O^{2-}$, Ho/$O^{2-}$, Yb/$O^{2-}$, Pr/$Cl^{-}$, Sm/$F^{-}$, Sm/$O^{2-}$ and Sm/$Cl^{-}$.

```
11
AgCo2F7Ominus1.xyz
  Ag      5.950540     5.991215     6.044541
  Co      8.415635     6.328347     4.112418
  Co      8.470004     5.679775     7.909705
  O       8.036140     6.002803     6.016673
  F       6.411167     5.611662     8.240284
  F       6.348086     6.373342     3.837018
  F       3.813023     5.980530     6.073304
  F       9.229692     4.966049     3.002974
  F       9.298935     7.049903     8.998327
  F       9.315320     4.033144     8.478564
  F       9.227435     7.983422     3.519711

11
AgCo2F8minus2.xyz
  Ag      6.017762     5.989951     6.052511
  Co      8.417012     6.363803     3.898178
  Co      8.455576     5.645806     8.109721
  F       6.441435     5.636763     8.115808
  F       6.386912     6.343913     3.983505
  F       3.994308     5.978378     6.096896
  F       9.118834     4.905856     2.967780
  F       8.223514     6.002397     6.005147
  F       9.171722     7.107824     9.017284
  F       9.184836     3.972047     8.485044
  F       9.104065     8.053454     3.501646

11
AgCo2F7Clminus2.xyz
  Ag      6.203718     5.991665     6.042627
  Co      8.324045     6.445441     3.431811
  Co      8.384661     5.563542     8.584443
  Cl      8.723137     6.003282     6.006562
  F       6.450220     5.635641     8.149059
  F       6.400720     6.351265     3.935020
  F       4.175980     5.980032     6.075063
  F       8.924032     4.999544     2.429354
  F       9.008799     7.009129     9.573009
  F       9.011028     3.872521     9.036848
  F       8.909636     8.148130     2.969724

11
AgCu2F7Ominus1.xyz
  Ag      5.857710     6.001231     6.041930
  Cu      8.411821     6.098476     4.151537
  Cu      8.456854     5.891922     7.874769
  F       6.395145     6.119674     8.199380
  F       6.330278     5.886507     3.871576
  F       3.717675     6.015984     6.073246
  F       8.798398     5.385488     2.434625
  O       7.920151     5.978842     6.017259
  F       8.876674     6.594727     9.588787
  F       9.913008     4.671071     7.804178
  F       9.838261     7.356267     4.176235

11
AgCu2F8minus2.xyz
  Ag      5.955950     6.001683     6.042652
  Cu      8.360001     6.142392     3.850354
  Cu      8.417739     5.861741     8.173254
  F       6.455326     6.158756     8.107900
  F       6.398708     5.844128     3.964068
  F       3.936497     6.002625     6.071078
  F       8.508814     5.408366     2.168897
  F       8.087870     6.000647     6.016499
  F       8.600991     6.561003     9.865515
  F       9.928519     4.856772     8.009729
  F       9.865558     7.162075     3.963578

11
AgNi2F7Ominus1.xyz
  Ag      5.991291     5.528459     6.048475
  Ni      8.383060     6.217710     4.094578
  Ni      8.451392     6.194617     7.924140
  F       6.533184     5.643794     8.276498
  F       6.446040     5.697227     3.804307
  F       3.927630     4.953372     6.081988
  F       8.774466     7.834202     3.109760
  O       8.007582     6.090118     6.015889
  F       8.832431     7.822291     8.890935
  F       9.648988     5.013086     8.874313
  F       9.519912     5.005314     3.112641

11
AgNi2F8minus2.xyz
  Ag      6.151895     5.574074     6.044552
  Ni      8.354744     6.378351     3.812214
  Ni      8.519709     6.045681     8.205243
  F       6.650593     5.383717     8.116413
  F       6.415287     5.985355     3.961493
  F       4.210902     5.021429     6.074324
  F       8.540828     7.853087     2.707176
  F       8.218192     6.163772     6.012599
  F       8.628481     7.847250     8.614454
  F       9.468322     4.871522     9.278187
  F       9.357024     4.875953     3.406868

11
AgNi2F7Clminus2.xyz
  Ag      6.433863     5.657481     6.043840
  Ni      8.270171     6.188024     3.138056
  Ni      8.377675     6.162580     8.881441
  Cl      8.821766     6.300706     6.001726
  F       6.629797     5.687637     8.200323
  F       6.547327     5.714884     3.880423
  F       4.479587     5.127329     6.077854
  F       8.247042     7.828315     2.316782
  F       8.390631     7.807087     9.693666
  F       9.220611     4.749195     9.691405
  F       9.097507     4.776954     2.308007
```

```
11
Ag-f06-Eu2F7O-minus1.xyz
  Ag     0.625216   -3.132445   -0.389102
  Eu     2.226206   -0.334432    0.405012
  Eu    -2.010194   -1.134117   -0.100889
  F     -3.018412   -0.808553    1.704828
  F      2.859795    1.075881   -1.006967
  F      1.033384   -5.029468   -0.811583
  F      2.768662   -2.419480   -0.053380
  O      0.183307   -1.082721    0.073021
  F     -2.770340    0.026868   -1.668771
  F     -1.645721   -3.250338   -0.592293
  F      2.599663    0.269454    2.374650
```

```
11
Ag-f07-Gd2F7O-minus1.xyz
  Ag     0.590992   -3.183169    0.008454
  Gd     2.229809   -0.312161    0.244909
  Gd    -1.984529   -1.115347   -0.268415
  F     -3.053627   -0.386777    1.369717
  F      2.959025    0.731033   -1.409565
  F      0.960170   -5.132902    0.023883
  F      2.737667   -2.437796    0.284527
  O      0.192110   -1.078525   -0.009189
  F     -2.647525   -0.326014   -2.083728
  F     -1.678960   -3.279433   -0.268125
  F      2.546433    0.701737    2.042055
```

```
11
Ag-f10-Ho2F7O-minus1.xyz
  Ag     0.588935   -3.186666    0.008003
  Ho     2.193445   -0.332192    0.237557
  Ho    -1.943191   -1.115086   -0.259251
  F     -2.996781   -0.374104    1.335724
  F      2.898155    0.719069   -1.375397
  F      0.955879   -5.135748    0.020638
  F      2.728393   -2.408285    0.269936
  O      0.193119   -1.086693   -0.004190
  F     -2.595243   -0.325589   -2.035306
  F     -1.685990   -3.243594   -0.260312
  F      2.514844    0.669536    1.997122
```

```
11
Ag-f13-Yb2F7O-minus1.xyz
  Ag     0.580290   -3.160457    0.008185
  Yb     2.168355   -0.344321    0.236332
  Yb    -1.914032   -1.108176   -0.259537
  F     -2.952264   -0.394290    1.315933
  F      2.883223    0.668759   -1.354954
  F      0.942323   -5.107832    0.023455
  F      2.726531   -2.373855    0.274360
  O      0.189018   -1.057965   -0.009048
  F     -2.555727   -0.357965   -2.018773
  F     -1.704643   -3.202876   -0.260672
  F      2.488490    0.619625    1.979243
```

```
11
Ag-f02-Pr2F7Cl-neutral.xyz
  Ag     0.564451   -3.056174    0.012784
  Pr     2.991985   -0.147337    0.335700
  Pr    -2.757973   -1.245243   -0.357482
  F     -4.051504   -0.866257    1.218440
  F      4.036170    0.648826   -1.269023
  F      0.934936   -5.000180    0.020592
  F      2.615016   -2.357462    0.260474
  Cl     0.061853   -0.420236    0.005373
  F     -3.618695   -0.807172   -2.192524
  F     -1.597927   -3.162574   -0.248085
  F      3.673252    0.594456    2.148275
```

```
11
Ag-f05-Sm2F7O-minus1.xyz
  Ag     0.663104   -3.075052   -0.626480
  Sm     2.208018   -0.359713    0.555899
  Sm    -2.019432   -1.168168   -0.052207
  F     -3.027709   -1.091399    1.799880
  F      2.777593    1.270309   -0.656401
  F      1.108048   -4.882528   -1.319458
  F      2.785476   -2.353713   -0.203230
  O      0.182709   -1.124876    0.119242
  F     -2.794754    0.191206   -1.467231
  F     -1.605940   -3.197645   -0.822787
  F      2.574451   -0.027775    2.607297
```

```
11
Ag-f05-Sm2F8-neutral.xyz
  Ag     0.635433   -3.434705    0.008438
  Sm     2.376139   -0.230876    0.258088
  Sm    -2.149482   -1.085587   -0.280068
  F     -3.141560   -0.375700    1.371004
  F      3.025007    0.775547   -1.409445
  F      1.000326   -5.381638    0.022732
  F      2.531515   -2.463020    0.244067
  F      0.145696   -0.833725   -0.003478
  F     -2.723888   -0.300311   -2.087719
  F     -1.482444   -3.220787   -0.242038
  F      2.634822    0.731449    2.052942
```

```
11
Ag-f05-Sm2F7Cl-neutral.xyz
  Ag     0.563335   -3.057030   -0.001130
  Sm     2.883711   -0.132442    0.305691
  Sm    -2.713277   -1.277201   -0.345365
  F     -3.887093   -0.814130    1.266266
  F      4.011922    0.617886   -1.220376
  F      0.954025   -4.995554    0.026833
  F      2.595378   -2.315633    0.238938
  Cl     0.016210   -0.413628   -0.046040
  F     -3.510702   -0.789317   -2.166236
  F     -1.600863   -3.171986   -0.239344
  F      3.538918    0.529680    2.115287
```

XYZ cartesian coordinates provided in angstroms for clusters with a neutral TEMPO ligand.

```
33
TEMPO_Gd2F5-plus1
  Gd     0.981224   -1.570426    1.920718
  Gd    -1.258610   -2.079848   -1.191899
  F     -0.217943   -2.926652    0.603399
  F      2.992633   -1.793190    1.882427
  F      0.120203   -0.582671    3.460688
  F     -3.277081   -2.002683   -1.105443
  F     -0.357114   -1.869202   -2.991837
  O     -0.020630   -0.297871    0.029731
  C     -1.084127    1.806870   -0.658124
  C      1.420966    1.671907   -0.250002
  C      0.959025    3.131456   -0.434401
  C     -0.363087    3.050920   -1.210984
  H      0.805209    3.597742    0.540228
  H      1.714168    3.712724   -0.961248
  H     -0.973830    3.943308   -1.081772
  H     -0.178275    2.929179   -2.279978
  N      0.097677    0.969306   -0.275604
  C     -1.892032    2.110432    0.617628
  H     -2.767369    2.700104    0.341734
  H     -2.239430    1.193548    1.095193
  H     -1.311803    2.680936    1.343028
  C     -1.945312    1.087469   -1.695562
  H     -2.575485    0.311038   -1.247722
  H     -2.649666    1.801316   -2.125814
  H     -1.350769    0.692955   -2.524366
  C      2.262717    1.164051   -1.434803
  H      3.231259    1.664445   -1.410371
  H      2.437502    0.089621   -1.369108
  H      1.790274    1.380263   -2.393444
  C      2.136789    1.429682    1.081326
  H      2.625467    0.450448    1.124101
  H      2.951511    2.149078    1.177460
  H      1.475690    1.581521    1.938628
```

```
33
TEMPO_Gd2F4O-neutral
  Gd     0.176791   -1.571689    2.232662
  Gd    -0.422691   -2.415712   -1.226461
  O      0.019832   -0.268693    0.055889
  N      0.120305    0.979002   -0.274396
  O     -0.013887   -2.998306    0.723524
  F     -1.536619   -1.103282    3.327440
  F      2.056611   -0.958692    2.922723
  F     -2.436465   -2.105145   -1.697834
  F      1.098165   -2.263059   -2.663790
  C     -1.088135    1.846054   -0.448195
  C      1.443550    1.641898   -0.509268
  C      1.012439    3.120215   -0.480316
  C     -0.438648    3.138747   -0.983601
  H      1.068821    3.495853    0.543589
  H      1.672202    3.734053   -1.092981
  H     -0.982508    4.019875   -0.644418
  H     -0.464736    3.137298   -2.074732
  C     -1.743085    2.036352    0.930214
  H     -2.631772    2.658770    0.815803
  H     -2.055029    1.082384    1.357748
  H     -1.069766    2.530863    1.632860
  C     -2.063177    1.207677   -1.439096
  H     -2.473552    0.265474   -1.076375
  H     -2.894084    1.896127   -1.603678
  H     -1.584351    1.028798   -2.404020
  C      1.965684    1.202921   -1.890225
  H      2.943026    1.660418   -2.053912
  H      2.080025    0.119561   -1.948653
  H      1.302671    1.521164   -2.696217
  C      2.438752    1.279663    0.589601
  H      2.709625    0.225365    0.570283
  H      3.348918    1.862568    0.436429
  H      2.054349    1.512716    1.583213
```

```
33
TEMPO_Ni2F5-minus1
  Ni    -0.336510   -1.163765    2.051294
  Ni     0.273402   -2.305475   -0.935963
  O     -0.185684   -0.273318   -0.123848
  N      0.007137    0.953399   -0.445424
  F      0.117318   -2.719907    0.984041
  F     -2.120600   -1.026508    2.431109
  F      0.923884   -0.544602    3.215719
  F     -0.778321   -3.388811   -1.955635
  F      1.997516   -1.903643   -1.474226
  C     -1.107564    1.935999   -0.340590
  C      1.361447    1.527599   -0.719643
  C      1.066677    3.028681   -0.523679
  C     -0.419110    3.203586   -0.878970
  H      1.240837    3.302811    0.519458
  H      1.719426    3.645872   -1.142986
  H     -0.850851    4.106364   -0.444423
  H     -0.546699    3.263266   -1.962840
  C     -1.536542    2.079693    1.129404
  H     -2.335420    2.822734    1.197907
  H     -1.914077    1.134046    1.524466
  H     -0.707202    2.407684    1.759263
  C     -2.283196    1.453480   -1.192248
  H     -2.648146    0.496914   -0.817372
  H     -3.099168    2.178523   -1.145050
  H     -1.987742    1.328049   -2.235958
  C      1.770179    1.196204   -2.165194
  H      2.744415    1.647438   -2.372941
  H      1.859083    0.114424   -2.283707
  H      1.050420    1.589664   -2.887657
  C      2.387357    0.960716    0.262625
  H      2.565949   -0.093416    0.048131
  H      3.328712    1.503990    0.140457
  H      2.056337    1.057545    1.298293
```

## 7. SI References